\documentstyle[12pt,epsf]{article}

\newcommand{\bfg}[1]{ \mbox{\boldmath $ #1 $} }
\newcommand{\dfrac}[2]{{\displaystyle\frac{#1}{#2}}}
\newcommand{\cfrac}[2]{\dfrac{\mathstrut #1}{#2}}

\begin{document}
\begin{flushright}
TIT/HEP-424/NP
\end{flushright}

\begin{center}
{\Large Weak Decay of $\Lambda$ in Nuclei : Quarks vs Mesons}
\end{center}

\begin{center}
Kenji Sasaki, Takashi Inoue, and Makoto Oka \\
        {\it Department of Physics, Tokyo Institute of Technology,}\\
        {\it Meguro, Tokyo 152-8551 Japan.}
\end{center}

\begin{abstract}
Decays of $\Lambda$ in nuclei, nonmesonic mode, 
are studied by using the $\Lambda N \to NN$ 
weak transition potential derived from
the meson exchange mechanism and the direct quark mechanism.
The decay rates are calculated both for the $\Lambda$ 
in symmetric nuclear matter and light hypernuclei.
We consider the exchange of six mesons 
($\pi, K, \eta, \rho, \omega, K^\ast$).
The form factor in the meson exchange mechanism 
and short range correlation are carefully studied.
\end{abstract}

\section{Introduction}
The $\Lambda$ particle is a neutral baryon 
with spin $1/2$ and strangeness $-1$.
It is unstable and has a finite lifetime of about $263$[ps].
The main decay modes of a free $\Lambda$ particle are the pionic ones,
\begin{eqnarray}
\Lambda & \to & p + \pi^- \nonumber \\
        & \to & n + \pi^0 \nonumber
\end{eqnarray}
where the final state nucleon has the momentum of about $100$[MeV/c].
This is one of the  $\Delta S =1$ non-leptonic weak interaction.
The important feature of this free $\Lambda$ decay 
is the empirical $\Delta I = 1/2$ rule.
The experimental branching ratio of two charge modes, 
$ \left( {\Gamma_{\pi^-}}/{\Gamma_{\pi^0}} \right)_{EXP} \simeq 1.78 $,
is close to the theoretical one 
with the assumption of $\Delta I = 1/2$ dominance, 
$ \left( {\Gamma_{\pi^-}}/{\Gamma_{\pi^0}} \right)_{Theory}^{\Delta I=1/2}=2$.
This fact tells us that 
the decay is dominated by $\Delta I = 1/2$ transition.
This $\Delta I = 1/2$ dominance is a prominent feature 
of all the observed $\Delta S =1$ non-leptonic weak interactions.
However, in the standard theory of electro-weak interaction, 
there are both $\Delta I= 1/2$ and $\Delta I= 3/2$ components.
Thus the effect of strong interaction should be significant 
for the explanation of enhancement of the $\Delta I= 1/2$ contribution
\cite{dgh:pre}.

\begin{figure}[t]
        \centerline{ \epsfxsize=6cm \epsfbox{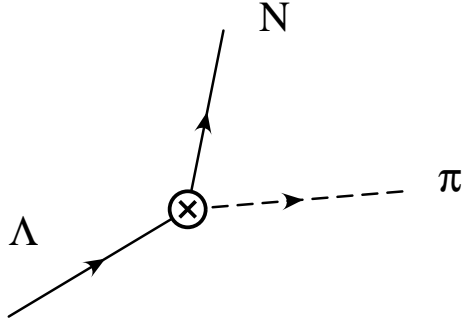} }
        \caption{Non-leptonic decay of hyperon.}
\end{figure}

The nucleus with the $\Lambda$ hyperon is called $\Lambda$ hypernucleus,
and is expressed $^A_\Lambda Z$ 
where $Z$ is the number of proton and $A$ is the mass number.
The $\Lambda$ hyperon in a nucleus 
falls into the ground state due to absence of the Pauli blocking effect 
and, therefore, is regarded as a probe of nuclear interior.
Since $\Lambda$ is not a stable particle, 
hypernucleus also decays via weak interaction,
which is so called weak decay of hypernucleus.
It is known that, when the $\Lambda$ hyperon is in the nuclear medium, 
the pionic decay mode is strongly suppressed 
and instead the nucleon induced decay modes,
\begin{eqnarray}
\Lambda + p &\to& n + p ~\ : ~\ {\mbox{\rm proton induced decay}} \\
\Lambda + n &\to& n + n ~\ : ~\ {\mbox{\rm neutron induced decay}}
\end{eqnarray}
become dominant.
This can be understood by considering
the Pauli principle for final state nucleons as for the mesonic decay
the momentum of the nucleon is less than the Fermi momentum in nuclear matter.
In the induced decay mode 
the momentum of the outgoing nucleons is about $400$[MeV/c] 
(assuming that the relative momentum of the initial $\Lambda$ and $N$ is zero).
It is much larger than the Fermi momentum, $k_F \simeq 270$[MeV/c].

The $\Lambda N \to NN$ transition is the $\Delta S=1$ reaction 
analogous to the weak $NN$ interaction or
the parity violating part of $NN$ force.
Although the weak $NN$ interaction is masked 
by the overwhelming strong interaction, 
the $\Lambda N \to NN$ transition is induced 
purely by the weak interaction, 
and therefore it gives us 
a unique chance to study the mechanism of baryon-baryon weak interaction.
At present, the direct observation of the $\Lambda N \to NN$ process 
is almost impossible because of the lack of hyperon beam or target.
Thus the weak decay of $\Lambda$ hypernucleus will be 
one of the clue of research for the $\Lambda N \to NN$ reaction.
Many theoretical and experimental efforts have been devoted 
to the nonmesonic decays of light and heavy hypernuclei~[2-20].
The mechanism of nonmesonic weak decay is still not clear, 
especially the theoretical prediction of the n/p ratio, 
which is the ratio of the neutron-induced decay to the proton-induced one, 
is not compatible with the experimental one.
This is because the theoretical value of proton-induced rate is 
much larger than the experimental one, 
which is measured to a good precision.
The value of the n/p ratio 
will be a key to understand the mechanism of the nonmesonic decay.

\begin{figure}[bt]
        \centerline{ \epsfxsize=6cm \epsfbox{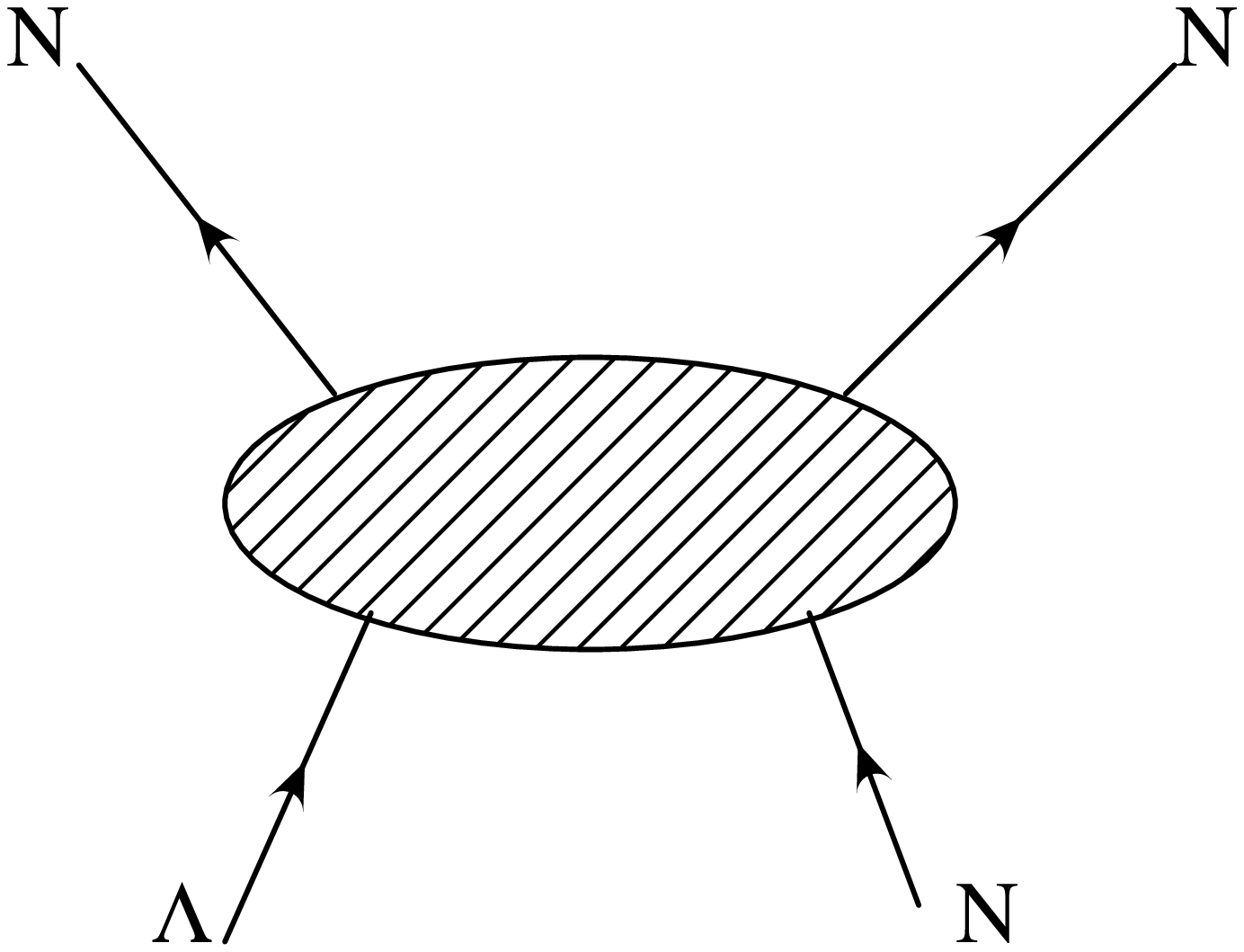} }
        \caption{$\Lambda N \to NN$ transition.} 
\end{figure}

A conventional picture of the two-baryon decay process, 
$\Lambda N\to NN$, is the one-pion exchange between the baryons,
where $\Lambda N\pi$ vertex is induced by the weak 
interaction~\cite{bld:prl,ost:npa,bmz:jmp,rmb:npa}.
In $\Lambda N\to NN$, the relative momentum of the final state
nucleon 
is about 400 MeV/c.
The nucleon-nucleon interaction at this momentum is dominated
by the short-range repulsion due to heavy meson exchanges and/or
to quark exchanges between the nucleons.
It is therefore expected that the short-distance interactions
will contribute to the two-body weak decay as well.
Exchanges of $K$, $\rho$, $\omega$, $K^*$ mesons
and also correlated two pions in the nonmesonic 
weak decays of hypernuclei have been
studied~\cite{mcg:prc,ttb:ptp,prb:prc,dub:anp,par:prc,shm:npa,ium:npa}
and it is found that the kaon 
exchange is significant, while the other mesons contribute 
less~\cite{par:prc}.

Several studies have been made on effects of quark 
substructure~\cite{chk:prc,mal:plb,iok:npa,iom:npa}.
In our recent analysis~\cite{iok:npa,iom:npa}, we employ an effective
weak Hamiltonian for quarks.
It was pointed out that the $\Delta I=1/2$ part of the Hamiltonian is
enhanced during downscaling of the renormalization point 
in the renormalization group equation. 
Yet a sizable $\Delta I=3/2$ component remains in the low energy
effective 
weak Hamiltonian.
We proposed to evaluate the effective Hamiltonian in the
six-quark wave 
functions of the two baryon systems and derived the ``direct quark''
weak transition potential for $\Lambda N \to NN$~\cite{iok:npa,iom:npa}.
Our analysis shows that the direct quark contribution
largely improves the discrepancy between the meson-exchange theory
and experimental data for the n/p ratios for light hypernuclei.
It is also found that the $\Delta I=3/2$ component of the effective Hamiltonian
gives a sizable contribution to $J=0$ transition amplitudes~\cite{mal:plb}.
Unfortunately, we cannot determine the $\Delta I=3/2$ amplitudes 
unambiguously from
the present experimental data~[17-20].

The aim of this paper is to combine the long-range meson exchange interactions
and the short-range direct quark transitions coherently in the study 
of $\Lambda$ decay in nuclear matter
and in light hypernuclei.
We first consider $\pi$ and $K$ exchanges 
and further introduce heavier mesons.
The direct quark mechanism represents the short-range part
of the transition and may therefore replace the heavy meson exchanges.
We compare the results of light + heavy meson exchanges 
with the results with light mesons + direct quark.
For the meson exchange, we particularly study 
effects of the form factors on the transition rates.
The pion exchange amplitudes are found to be sensitive 
to the choice of the form factors.
We propose to use a soft pion-baryon form factors
so that the strong tensor transition is suppressed.
The direct quark mechanism yields a large neutron-induced amplitude
and therefore improves the n/p ratio.
We, however, find that its enhancement is not large
enough to explain the observed large n/p ratio.

This paper is organized as follows.
In section 2,
we present the transition potential for the $\Lambda N \to NN$ transition.
In section 3 and 4,
the transition potential is applied to the decay 
in the nuclear matter and light hypernuclei, respectively.
Discussions and conclusion are given in section 5.

\section{Theoretical approach to $\Lambda N \to NN$ }
Since the $\Lambda N \to NN$ transition involves a large energy transfer, 
the short range reactions may contribute significantly.
Thus we expect that the substructure of the baryons, or
the quark degree of freedom play important roles in this transition
\cite{chk:prc,mal:plb,iok:npa,iom:npa}.
In this paper 
we assume the hybrid mechanism 
to describe $\Lambda N \to NN$ decay, 
meson exchange mechanism (ME) and direct quark mechanism (DQ).
Such combination has been successfully employed 
in describing the strong baryon-baryon interactions.

\subsection{Meson Exchange Mechanism}
\begin{figure}[tb]
\centerline{ \epsfxsize=6cm \epsfbox{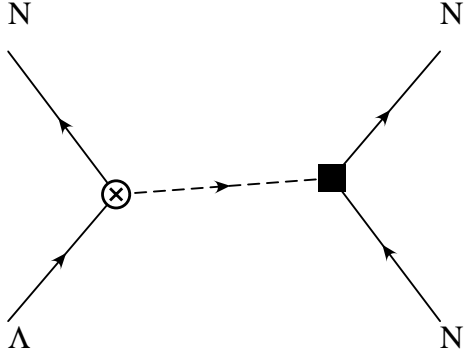} }
\caption{Meson exchange diagram for $\Lambda N \to NN$ decay.
         $\otimes$ denotes the weak vertex for the nonstrange mesons
         and the strong vertex for the strange mesons.}
\label{obe}
\end{figure}

The one pion exchange (OPE) process, 
where the emitted virtual pion from weak $\Lambda \to N \pi$ vertex 
is absorbed by nucleon in the nucleus,
shown in Fig.~\ref{obe}, is studied to describe $\Lambda N \to NN$ decay
\cite{bld:prl,ost:npa,bmz:jmp,rmb:npa}.
This mechanism provides the simplest but indispensable picture
because it is based on the main decay mode of the free $\Lambda$.
This mechanism yields a long range transition potential,
which is obtained by evaluating
the diagram using the following vertices,
\begin{eqnarray}
{\cal{H}}_s^\pi
    & = & 
         ig_{\mbox{\tiny $NN \pi$}} \overline{\psi}_N \gamma_5 
         \vec{\tau} \cdot \vec{\phi}_\pi \psi_N
        \label{eqn:phenostrong}
 \\
{\cal{H}}_w^{\pi} 
    & = & 
         iG_Fm_{\pi}^2 \overline{\psi}_N ( A_\pi + B_\pi \gamma_5 ) 
         \vec{\tau} \cdot \vec{\phi}_\pi \psi_\Lambda.
       \label{eqn:phenoweak}
\end{eqnarray}
Both the strong coupling constant, $g_{\mbox{\tiny $NN \pi$}}$,
and the weak ones, $A_\pi$ and $B_\pi$,
are well determined from experiment.
Their values are listed in Table~\ref{cct}.
It is known that the OPE transition potential enhances
the $J=1$ transition due to the strong tensor term
and therefore the amplitude of proton-induced mode is much larger 
than neutron-induced one.
This tensor dominance causes the n/p ratio problem, namely,
the observed n/p ratio ($\simeq 1$) can not be explained
by the OPE transition.

While the OPE is significant
for the long-distance baryon-baryon interaction, 
the short range reaction mechanism is also important 
in $\Lambda N \to NN$
due to the large energy transfer involved.
Within the meson exchange model,
a shorter range contributions 
may come from the exchange of the heavier mesons
\cite{mcg:prc,ttb:ptp,prb:prc,dub:anp,par:prc}
and correlated two pions meson~\cite{shm:npa,ium:npa}.
In ref \cite{par:prc}
the authors consider the exchange of all the octet pseudoscalar 
and vector mesons, 
$\pi$, $K$, $\eta$, $\rho$, $\omega$, and $K^\ast$.
Although the $\Lambda N \pi$ weak coupling constant 
is determined phenomenologically, 
all the other weak couplings in ref \cite{par:prc}
are estimated theoretically by assuming
the $SU(6)_w$ symmetry \cite{dub:anp}.
While for the strong vertices, 
the couplings are taken from the Nijmegen YN potential model (soft core)
\cite{nrs:prd}.

The strong vertices are given by
\begin{eqnarray}
    & & {\cal{H}}_s^K 
         =  ig_{\mbox {\tiny $\Lambda N K$}} \overline{\psi}_N \gamma_5 
            \phi_K \psi_\Lambda 
            \\
    & & {\cal{H}}_s^\eta 
         =  ig_{\mbox{\tiny $NN \eta$}} \overline{\psi}_N \gamma_5 
            \phi_\eta \psi_N 
            \\
    & & {\cal{H}}_s^\rho 
         =  \overline{\psi}_N \left(g_{\mbox{\tiny $NN \rho$}}^V \gamma^\mu 
            + i \cfrac{g_{\mbox{\tiny $NN \rho$}}^T}{2M} \sigma^{\mu \nu} 
            q_\nu \right) \vec{\tau} \cdot \vec{\phi}^\rho_\mu \psi_N 
            \\
    & & {\cal{H}}_s^{K^\ast} 
         =  \overline{\psi}_N \left(g_{\mbox{\tiny $\Lambda N K^\ast$}}^V 
            \gamma^\mu + i \cfrac{g_{\mbox{\tiny $\Lambda N K^\ast$}}^T}{2M} 
            \sigma^{\mu \nu} q_\nu \right) \phi^{K^\ast}_\mu \psi_\Lambda 
            \\
    & & {\cal{H}}_s^\omega 
         =  \overline{\psi}_N \left(g_{\mbox{\tiny $NN \omega$}}^V \gamma^\mu 
            + i \cfrac{g_{\mbox{\tiny $NN \omega$}}^T}{2M} \sigma^{\mu \nu} 
            q_\nu \right) \phi^\omega_\mu \psi_N 
\end{eqnarray}
while the weak vertices are parameterized as
\begin{eqnarray}
    & & {\cal{H}}_w^K 
         =  iG_Fm_\pi^2 \left[ (\overline{\psi}_N)_s
            (C_K^{PV} + C_K^{PC} \gamma_5) 
            \phi_K^\dagger \psi_N + \overline{\psi}_N 
            (D_K^{PC} + D_K^{PV} \gamma_5) \psi_N 
            (\phi_K^\dagger)_s \right] 
            \\
    & & {\cal{H}}_w^\eta 
         =  iG_Fm_\pi^2 \overline{\psi}_N 
            \left(A_\eta + B_\eta \gamma_5 \right) 
            \phi_\eta \psi_\Lambda \\
    & & {\cal{H}}_w^\rho 
         =  G_Fm_{\pi}^2 \overline{\psi}_N \left( \alpha_\rho \gamma^\mu 
            - \beta_\rho i \cfrac{\sigma^{\mu \nu} q_\nu}{2 \overline{M}} 
            + \epsilon_\rho \gamma^\mu \gamma_5 \right) 
            \vec{\tau} \cdot \vec{\psi}^\rho_\mu \psi_\Lambda 
            \\
    & & {\cal{H}}_w^{K^\ast} 
         =  G_Fm_{\pi}^2 \left( [C_{K^\ast}^{PC,V} (\overline{\psi}_N)_s 
            {\phi^{K^\ast}_\mu}^\dagger \gamma^\mu \psi_N 
            + D_{K^\ast}^{PC,V} \overline{\psi}_N 
            \gamma^\mu \psi_N ({\phi^{K^\ast}_\mu}^\dagger)_s] \right. 
            \nonumber \\
    & &  \verb|           | 
         -  i [C_{K^\ast}^{PC,V} (\overline{\psi}_N)_s {K^\ast_\mu}^\dagger 
            \frac{\sigma^{\mu \nu} q_\nu}{2M} N + D_{K^\ast}^{PC,V} 
            \overline{\psi}_N \frac{\sigma^{\mu \nu} q_\nu}{2M} \psi_N 
            ({\phi^{K^\ast}_\mu}^\dagger)_s] 
            \nonumber \\
    & &  \verb|           | 
            \left. + [C_{K^\ast}^{PV} (\overline{\psi}_N) 
            {\phi^{K^\ast}_\mu}^\dagger \gamma^\mu \gamma_5 \psi_N 
         +  D_{K^\ast}^{PV} 
            \overline{\psi}_N \gamma^\mu \gamma_5 \psi_N 
            ({\phi^{K^\ast}_\mu}^\dagger)_s] 
            \right) 
            \\
    & & {\cal{H}}_w^\omega 
         =  G_Fm_{\pi}^2 \overline{\psi}_N \left( \alpha_\omega \gamma^\mu 
         -  \beta_\omega i \cfrac{\sigma^{\mu \nu} q_\nu}{2 \overline{M}} 
         +  \epsilon_\omega \gamma^\mu \gamma_5 \right) 
            \phi^\omega_\mu \psi_\Lambda.
\end{eqnarray}
We choose a convention that the three-momentum transfer
${\bf{q}}$ is directed towards to the strong vertex, 
and we assign the spurious isospin state to the $\Lambda$ field 
which behaves like an isospin $|\frac{1}{2} -\frac{1}{2} \rangle$.
(The same isospurion $\left( 0 \atop 1 \right)$ is also employed 
to $\psi^N_{s}(=n)$ and $\phi^K_{s}(=K^0)$.)
The value of these couplings are given in Table~\ref{cct}.
We do not present the explicit form of the potential here,
because it is given in detail in ref \cite{par:prc}.

The potential is given by the Fourier transformation of the product 
of the propagator and the vertices in the Breit-Fermi reduction.
\begin{table}[tb]
\caption{The strong and weak meson baryon coupling constants.
         The strong coupling constants are taken according to the 
         Nijmegen potential (soft core)~\cite{nrs:prd}.
         The weak coupling constants are estimated by ref~\cite{par:prc}.
         They are in units of $G_F m_\pi^2 =2.21 \times 10^{-7}$.
         The cutoff parameters for the DP form factor
         are also given.}
\begin{center}
\begin{tabular}{|c||l|l|l|r|r|}
\hline
Meson  & Strong c.c. & \multicolumn{2}{c|}{Weak c.c.} & \multicolumn{2}{c|}{$\Lambda$[MeV]} \\ 
\cline{3-6}
 & & \multicolumn{1}{c|}{PC} & \multicolumn{1}{c|}{PV} & soft & hard \\
\hline
$\pi$  & $g_{NN\pi}=13.3$  & $B_\pi=-7.15$  & $A_\pi=1.05$  & 800  & 1300 \\
       & $g_{\Lambda \Sigma \pi}=12.0$ &    &               &      &      \\
\hline
$\eta$ & $g_{NN\eta}=6.40$ & $B_\eta=-14.3$ & $A_\eta=1.80$ & 1300 & 1300 \\
       & $g_{\Lambda \Lambda \eta}=-6.56$ & &               &      &      \\
\hline
$K$    & $g_{\Lambda NK}=-14.1$ & $C_K^{PC}=-18.9$  & $C_K^{PV}=0.76$  & 1200 & 1200 \\
       & $g_{N \Sigma K}=4.28$  & $D_K^{PC}=6.63$   & $D_K^{PV}=2.09$ & &  \\
\hline
$\rho$ & $g_{NN \rho}^V=3.16$ & $\alpha_\rho=-3.50$ & $\epsilon_\rho=1.09$ & 1400 & 1400 \\
       & $g_{NN \rho}^T=13.1$ & $\beta_\rho=-6.11$  & & & \\
       & $g_{\Lambda \Sigma \rho}^V=0$    &         & & & \\
       & $g_{\Lambda \Sigma \rho}^T=11.2$ &         & & & \\
\hline
$\omega$ & $g_{NN \omega}^V=10.5$ & $\alpha_\omega=-3.69$ & $\epsilon_\omega=-1.33$ & 1500 & 1500 \\
         & $g_{NN \omega}^T=3.22$ & $\beta_\omega=-8.04$  & & & \\
         & $g_{\Lambda \Lambda \omega}^V=7.11$          & & & & \\
         & $g_{\Lambda \Lambda \omega}^T=-4.04$         & & & & \\
\hline
$K^\ast$ & $g_{\Lambda N K^\ast}^V=-5.47$ & $C_{K^\ast}^{PC,V}=-3.61$ & $C_{K^\ast}^{PV}=-4.48$ & 2200 & 2200 \\
         & $g_{\Lambda N K^\ast}^T=-11.9$ & $C_{K^\ast}^{PC,T}=-17.9$ & & & \\
         & $g_{N \Sigma K^\ast}^V=-3.16$  & $D_{K^\ast}^{PC,V}=-4.89$ & $D_{K^\ast}^{PV}=0.60$ & & \\
         & $g_{N \Sigma K^\ast}^V=-3.16$  & $D_{K^\ast}^{PC,T}=9.30$  & & & \\
\hline
\end{tabular}
\end{center}
\label{cct}
\end{table}
As the baryons and also the mesons have finite sizes and structures,
we need to take a form factor at each meson-baryon vertex into account
\cite{vgj:prc}.
If we assume the same form factor $F({\bf q}^2)$ 
for both the strong and weak vertices,
the potential is given as
\begin{equation}
V ({\bf{r}}) 
    = \int \frac{d^3q}{(2\pi)^3} \frac{e^{i {\bf{q \cdot r}}}}
   {{\bf{q}}^2 + \mu^2 - q_0^2} {\cal{O}} ({\bf{q}}) F^2({\bf q}^2)
   \label{ffp}
\end{equation}
where ${\cal O}$ is a product of the vertex operators 
and the coupling constants.
The operator ${\cal O}$ includes the isospin operator,
$$ \langle I \mid {\bfg{\tau}}_1 \cdot {\bfg{\tau}}_2 \mid I \rangle 
     = \left\{ \begin{array}{cl}
       -3 & I=0 \\
        1 & I=1 
       \end{array} \right. .
$$
In this case 
we use the initial state which is antisymmetrized 
in the flavor space.

\begin{figure}[tb]
\centerline{ \epsfxsize=12cm \epsfbox{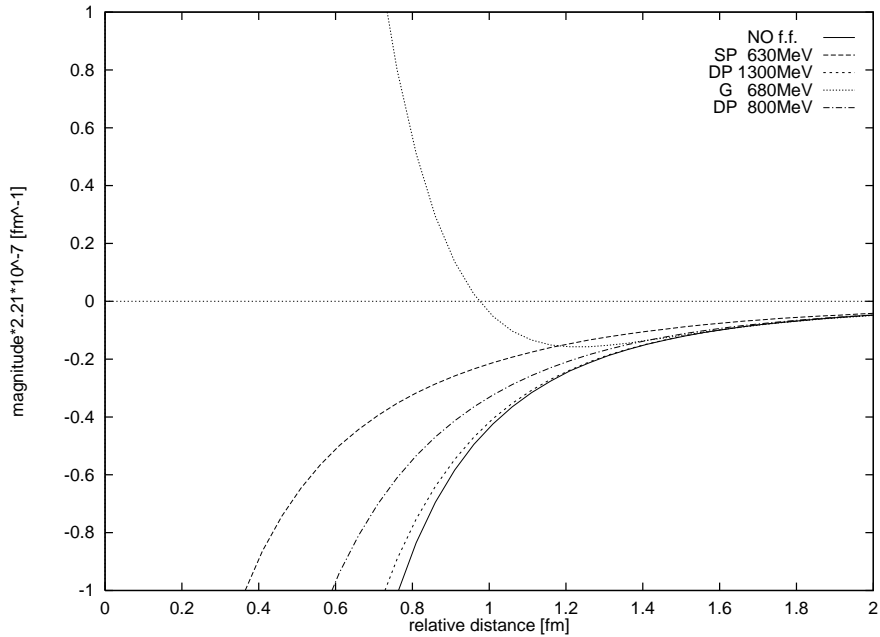} }
\caption{The OPE potentials for the $^3S_1 \to ^3D_1$ for various form factors.}
\label{pot9}
\end{figure}

A standard choice of $F^2({\bf q}^2)$ is 
the square of the monopole form factor,
\begin{equation}
F^2_{DP}({\bf{q}}^2) 
    = \left( \cfrac{\Lambda_{DP}^2 - \mu^2}{\Lambda_{DP}^2 
         + {\bf{q}}^2} \right)^2
\end{equation}
which we call ``double  pole'' (DP) form factor.
While in literatures a simple form
\begin{equation}
F^2_{SP}({\bf{q}}^2) 
    = \cfrac{\Lambda_{SP}^2 - \mu^2}{\Lambda_{SP}^2 + {\bf{q}}^2}
\end{equation}
is often used, which we call ``single pole'' (SP) form factor.
The cutoff parameter $\Lambda$ 
is chosen for individual mesons independently.
These form factors are normalized at the on-mass-shell point as
$F^2_{DP}(-\mu^2) = F^2_{SP}(-\mu^2) = 1$.
There is another type of form factor, ``Gaussian'' (G),
\begin{equation}
F^2_G({\bf{q}}^2) 
    = \exp \left( -\frac{{\bf{q}}^2}{\Lambda^2} \right),
\end{equation}
which has the advantage of the consistency 
with the quark structure for the baryon, 
if the cutoff parameter is taken according to 
the size of quark distribution in the baryon.
This form factor is normalized as $F^2_G(0)=1$.
It should be noted that 
the SP and DP form factors give similar effects.
Near at the mass shell point, $F^2_{SP}$ and $F^2_{DP}$ are expanded as
\begin{equation}
F^{SP} = \frac{\Lambda_{SP}^2-\mu^2}{\Lambda_{SP}^2 +
         {\bf{q}}^2} = 1 - \left( \frac{{\bf{q}}^2 + \mu^2}
                                        {\Lambda_{SP}^2 + {\bf{q}}^2} \right)
\end{equation}
\begin{equation}
F^{DP} = \left( \frac{\Lambda_{DP}^2-\mu^2}{\Lambda_{DP}^2 + 
        {\bf{q}}^2} \right)^2 \simeq 1 - 2 \left( \frac{{\bf{q}}^2+ \mu^2}
                                        {\Lambda_{DP}^2 + {\bf{q}}^2} \right)
\end{equation}
Therefore, the SP and DP form factors are identical
if we set $ \Lambda_{DP} \simeq \sqrt{2} \Lambda_{SP}$ 
by assuming $\Lambda^2 \gg \mu^2$.
On the other hand, 
the G form factor behaves differently at short distances.
Fig.~\ref{pot9} shows the behaviors of the potential 
for several form factors 
in the $\Lambda N$ : $^3S_1$ - $NN$ : $^3D_1$ OPE transition 
at the relative momentum of $k_r = 1.97$fm$^{-1}$.
As can be seen clearly, the potential has a node for the G form factor, 
while the others do not behave like that.
This oscillation suppresses the tensor transition 
although the long distance behavior is similar to the `` hard '' DP form factor 
or that without the form factor.
We test these three form factors and compare the results.

We cannot neglect the finite energy transfer $q_0$ in eq. (\ref{ffp}).
In the case of typical nucleon-nucleon interaction, 
energy transfer is small and negligible, 
but in the case of $\Lambda N \to NN$ 
this energy transfer is not negligible because
the mass-difference between $\Lambda$ and $N$ causes a large $q_0$.
If we assume that 
the initial $\Lambda$ - $N$ pair has a low kinetic energy, 
this energy transfer is regarded as a constant, 
$
q_0 \simeq (m_\Lambda - m_N)/2 \simeq 88{\mbox{\rm MeV}}.
$
Accordingly we 
introduce effective masses of the exchanged mesons,
$
\tilde{\mu}_i = \sqrt{\mu_i^2 - (m_\Lambda - m_N)^2/4}
$
and perform following replacement,
\begin{equation}
\frac{1}{{\bf{q}}^2 + \mu^2 - q_0^2} \to \frac{1}{{\bf{q}}^2 + \tilde{\mu}^2}.
\end{equation}
This effect leads to measurable changes.
For example, the pion mass is reduced by about $25\%$,
and the range of OPE becomes longer.

\subsection{Direct Quark Mechanism}

\begin{figure}[tb]
\centerline{ \epsfxsize=6cm \epsfbox{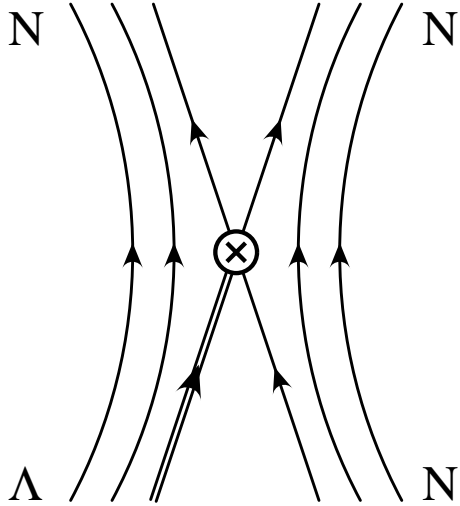} }
\caption{Direct quark mechanism for non-mesonic decay of hyperon.
The double line indicates the strange quark, and $\otimes$ stands for 
the weak vertex.}
\label{dqfig}
\end{figure}

Studies of nuclear forces in the quark cluster model have revealed 
that the short range repulsion between nucleons is explained in terms 
of the quark exchange interactions~\cite{QCM}.
The range of the strong 
repulsion is about 0.5 fm and corresponds to a momentum transfer of 
about 400 MeV/c.  The same momentum transfer is required in the 
$\Lambda N\to NN$ transition in order to satisfy the energy-momentum 
conservation.
It is therefore expected that the quark substructure of the baryons 
plays a significant role also in the short range part of 
the $\Lambda N \to NN$ interaction.
Recently, we proposed the direct quark (DQ) mechanism, in which
the weak interaction between quarks in baryons causes the decay
without exchanging mesons Fig.~\ref{dqfig}~\cite{iok:npa}.

The weak interaction among constituent quarks is described as 
a $\Delta S= 1$ effective weak Hamiltonian,
which consists of various four quark weak vertices
derived in the renormalization group approach to include QCD 
corrections to the $su\to ud$ transition mediated 
by the $W$ boson~\cite{psw:npb}.
Such effective weak Hamiltonian has been applied to the decays of 
kaons and hyperons with considerable success.  Our aim is to explain 
the short range part of the weak baryonic interaction by using the 
same interaction so that we are able to confirm the consistency 
between the free hyperon decays and decays of hypernuclei.

The transition potential is calculated by evaluating the effective 
Hamiltonian in the nonrelativistic valence quark model~\cite{iok:npa}.
The valence quark wave function of two baryon system is taken from 
the quark cluster model, which takes into account the 
antisymmetrization among the quarks.
Then the obtained transition potential has been applied to 
the weak decay of light hypernuclei~\cite{iom:npa}.

In ref~\cite{iok:npa}, the DQ transition potential is calculated 
for all the two-body channels with the initial 
$\Lambda N (L=0)$ and the final $NN (L'=0,1)$ states.
The explicit forms of the transition potential are given in the 
momentum space representation. 
In the present analysis, we employ the coordinate space formulation, 
so that realistic nuclear wave functions with short range 
correlations can be easily handled.
The DQ transition potentials in the coordinate space contain
nonlocal terms as a result of the quark antisymmetrization.
It also has terms with a derivative operator acting on the initial 
relative coordinate.
Thus the general form of the transition potential is 
\begin{eqnarray}
{V_{DQ}}^{L L'}_{SS'J}(r,r')
     &=& 
           \langle NN: L' S' J|V({\vec r'},
           {\vec r}) |\Lambda N : L S J\rangle \nonumber \\
     &=& 
           V_{loc}(r)\, {\delta(r-r')\over r^{2}}
         + V_{der}(r)\, {\delta(r-r')\over r^{2}} \partial_{r}
         + V_{nonloc} (r,r')
\label{dqpot}
\end{eqnarray}
where $r$ ($r'$) denotes the relative radial coordinate of the 
initial (final) two-baryon state, and $\partial_r$ stands for the 
derivative of the initial $\Lambda N$ relative wave function.
The explicit forms of  $V_{loc}(r)$, $V_{der}(r)$, and 
$V_{nonloc}(r,r')$, are given in Appendix.
It should be noted here that the DQ potential does not contain the 
tensor terms as we truncated the nonrelativistic expansion at the $O(p^2)$.  
We expect that the tensor component of the DQ transition 
is negligibly small.

Combining the direct quark (DQ) and meson exchange (ME) potentials, 
we obtain the total transition potential as
\begin{equation}
V(r,r') = V^{ME}(r) {\delta(r-r')\over r^2} + V^{DQ}(r,r')
\end{equation}
The relative phase between ME and DQ is fixed so that the weak quark 
Hamiltonian gives the correct amplitude of $\Lambda \to N\pi$ 
transition~\cite{iom:npa}.  
Note that the relative phases among various meson exchange potentials 
are determined according to the $SU(6)_w$ symmetry.

\section{Nuclear Matter Calculation}
Here we study the $\Lambda N \to NN$ decay in nuclear matter 
in order to investigate our approach to $\Lambda N \to NN$
with minimizing wave function model ambiguity.
On top of that, the recent experimental data 
of life-times of heavy hypernuclei indicate 
that the nonmesonic decay rate of $\Lambda$ in heavy hypernuclei 
is only about $20\%$ larger than the free $\Lambda$ decay rate~\cite{bmz:jmp}.
This saturation suggests 
the short-range nature of the $\Lambda N \to NN$ decay.
We study the $\Lambda$ decay in nuclear matter
as an approximation to the heavy $\Lambda$ hypernuclei.

\begin{table}[tb]
\caption{Possible $^{2S+1}L_J$ channels 
         for the $\Lambda N \to NN$ transitions. 
         $I_f$ stands for the total isospin of the final state. 
         PC and PV indicate the parity conserving and 
         parity violating channels respectively.}
\begin{center}
\begin{tabular}{ccccccccc}
 $^1S_0$ & $\to$ & $^1S_0$ & : & $I_f=1$ & $a_p$ & $a_n$ &:& PC \\
 $^1S_0$ & $\to$ & $^3P_0$ & : & $I_f=1$ & $b_p$ & $b_n$ &:& PV \\
 $^3S_1$ & $\to$ & $^3S_1$ & : & $I_f=0$ & $c_p$ &       &:& PC \\
 $^3S_1$ & $\to$ & $^3D_1$ & : & $I_f=0$ & $d_p$ &       &:& PC \\
 $^3S_1$ & $\to$ & $^1P_1$ & : & $I_f=0$ & $e_p$ &       &:& PV \\
 $^3S_1$ & $\to$ & $^3P_1$ & : & $I_f=1$ & $f_p$ & $f_n$ &:& PV \\
\end{tabular}
\end{center}
\label{cha}
\end{table}

We assume that nuclear matter is symmetric ($N_n = N_p$) 
and the $\Lambda$ is at rest in it.
In this assumption 
the initial state density is given as 
\begin{equation}
\int^{k_F}_0 \frac{d^3k}{(2 \pi)^3} \sum_{S_N} 
     \times \frac{1}{2} \sum_{S_\Lambda} \times \sum_{p,n}
\end{equation}
where $S_N$ and $S_\Lambda$ are the spin of nucleon and $\Lambda$
hyperon respectively.
Similarly, the final state density is 
\begin{equation}
\int \frac{d^3k_1}{(2 \pi)^3} \frac{d^3k_2}{(2 \pi)^3}
     (2 \pi)^4 \delta^4(E.M.C.) \sum_{S_{N_1} S_{N_2}} 
\end{equation}
where the $\delta(E.M.C)$ stands for the $\delta$ function 
coming from the energy momentum conservation.
Then the $\Lambda N \to NN$ decay rate in symmetric nuclear matter 
becomes
\begin{equation}
\Gamma_{NM} =
   \int_0^{k_F} \frac{d^3 k  }{(2 \pi)^3} 
   \int         \frac{d^3 k_1}{(2 \pi)^3} 
   \int         \frac{d^3 k_2}{(2 \pi)^3} (2 \pi)^4 \delta^4(E.M.C) 
   \frac{1}{2} \sum_{spin} \sum_{p,n} \mid 
         \langle \Psi_f(k_1,k_2) \mid V  \mid \Psi_i(k,0) \rangle 
         \mid^2
\end{equation}
where $\mid \Psi_i (k,0) \rangle$ stands for 
an initial $\Lambda N$ state with the momentum of the nucleon $k$, 
and $\mid \Psi_f (k_1, k_2) \rangle$ stands for the final two nucleon state 
with the momenta $k_1$ and $k_2$ respectively.
The $V$ is $\Lambda N \to NN$ transition potential.

We employ the plane wave with the short range correlation
for the wave function of both the initial and final states. 
The short range correlation represents the short range repulsion 
between two baryons.
Therefore the configuration space wave function for the two baryon system
with total momentum ${\bf{K}}$ and relative momentum ${\bf{k}}$ is 
\begin{equation}
\langle {\bf R},{\bf r} \mid \Psi({\bf K},{\bf k}) \rangle 
     = \frac{1}{\sqrt 2} e^{i {\bf{K \cdot R}}}
       [ e^{i {\bf{k \cdot r}}} - (-1)^{S+I} e^{-i {\bf{k \cdot r}}}] 
           \chi_{S m_S} \chi_{I I_3}  \times f (r)
\end{equation}
where $f(r)$ is the correlation function 
and $S$ and $I$ are the spin and isospin 
of the two baryon system, respectively. 
We apply the correlation function proposed in ref~\cite{par:prc}.
For the initial $\Lambda N$, we employ
\begin{equation}
f_i(r) = \left( 1-e^{-\frac{r^2}{a^2}} \right)^n + br^2 e^{-\frac{r^2}{c^2}}
\end{equation}
where $a=0.5$[fm], $b=0.25$[fm$^{-2}$], $c=1.28$[fm], 
$n=2$, and for the final $NN$
\begin{equation}
f_f(r) = 1-j_0( q_c r ) 
\end{equation}
where $q_c=3.93$ fm$^{-1}$.
The initial state correlation is obtained 
from a macroscopic finite nucleus G-matrix calculation, 
and the final state correlation 
gives a good description of nucleon pairs in $^4$He.

In the present calculation, we only consider the relative $L = 0$ 
for the initial $\Lambda - N$ system.
Then the possible transition channels 
are those given in Table~\ref{cha}.
The decay rate is decomposed as
\begin{equation}
\Gamma_{NM}
    = 
       \Gamma_p + \Gamma_n 
\end{equation}
where the $\Gamma_p$ and $\Gamma_n$ stand for the proton- 
and neutron-induced decay rates respectively, 
and are given as
\begin{eqnarray}
& & \Gamma_p 
       = 
       \frac{m_N}{(2\pi)^5 \mu^3} \int^{\mu k_F}_0 k^2 k_0 dk \frac{1}{4} 
       \left( |a_p|^2 + |b_p|^2 + 3|c_p|^2 + 3|d_p|^2 + 3|e_p|^2 + 3|f_p|^2 
       \right) \\
& & \Gamma_n 
       = 
       \frac{m_N}{(2\pi)^5 \mu^3} \int^{\mu k_F}_0 k^2 k_0 dk \frac{1}{4} 
       \left( |a_n|^2 + |b_n|^2 + 3|f_n|^2 
       \right) 
\end{eqnarray}
\begin{equation}
k_0^2 \equiv m_N ( m_\Lambda - m_N ) + \frac{k^2}{2\mu} \nonumber
\end{equation}
\begin{equation}
\mu \equiv \frac{m_\Lambda}{m_N + m_\Lambda} \nonumber
\end{equation}
where the $a$, $b$, $\cdots$, $f$ are the matrix elements of 
the transition potential, for example,
\begin{equation}
a_p \equiv \langle ^1S_0: [pn]_{I=1} \mid V 
                                  \mid {^1S_0}: [\Lambda p]_s \rangle
\end{equation}

\subsection{Form Factor Dependence}
As the $\Lambda N \to NN$ scattering involves a large momentum transfer, 
$\sqrt{m_N(m_\Lambda - m_N)c^2} \sim 400MeV/c$, 
the amplitudes are rather sensitive to the form factor.
Here we show the dependence on form factor 
in meson exchange potential.
First, we examine the form factor of OPE potential.
Table~\ref{ffc} shows the calculated decay rate
when we employ only OPE induced $\Lambda N \to NN$ potential.
The SP form has often been used in literatures \cite{mcg:prc,ttb:ptp}.
McKeller and Gibson employed the SP form factor 
with $\Lambda_{SP}^2 = 20m_\pi^2$, or $\Lambda_{SP} \sim 630$MeV.
This form factor is based on the dispersion relation analysis 
with the semi-pole approximation to the $\pi NN$ form factor.
It is a very soft form factor, 
which cuts off the short-range part rather drastically
and the resulting decay rate becomes small.
The tensor transition, $\Lambda p$ : $^3S_1 \to pn$ : $^3D_1$, 
is most affected by such form factor. 
We find that the tensor transition amplitudes 
are reduced by a factor two or more by the form factor.
Therefore total decay rate much reduced as shown in Table~\ref{ffc}

\begin{table}[tb]
\caption{Decay rates of $\Lambda$ in nuclear matter 
         (in units of $\Gamma_\Lambda = (263 \times 10^{-12}sec)^{-1}$)
         for various choices of the coupling form factors
         in the OPE mechanism.}
\begin{center}
\begin{tabular}{||c||ll|c|c|c|c|c||} 
\hline
 & & & $total$ & $\Gamma_p$ & $\Gamma_n$ & ${\Gamma_n}/{\Gamma_p}$ & $PV/PC$ \\
\hline
$\pi$ & no-f.f.&                    & 2.819 & 2.571 & 0.248 & 0.097 & 0.358 \\
      & SP & $\Lambda_\pi = 630$MeV & 1.103 & 0.989 & 0.114 & 0.116 & 0.408 \\
      &    & $\Lambda_\pi = 920$MeV & 2.332 & 2.116 & 0.216 & 0.102 & 0.376 \\
      & DP & $\Lambda_\pi =1300$MeV & 2.575 & 2.354 & 0.221 & 0.094 & 0.337 \\
      &    & $\Lambda_\pi = 800$MeV & 1.850 & 1.702 & 0.148 & 0.087 & 0.282 \\
      & G  & $\Lambda_\pi = 680$MeV & 1.514 & 1.129 & 0.386 & 0.342 & 0.580 \\
 \hline
\end{tabular}
\end{center}
\label{ffc}
\end{table}

The DP form factor is popular in meson exchange potential models 
of the nuclear force.
The Bonn potential~\cite{mhe:pre}, for instance, employs the DP 
with $\Lambda_{DP}(\pi) = 1300$MeV.
This form factor is extremely hard so that 
the short range part of the meson exchange potential becomes relevant.
The DP form factor is regarded as the product of the monopole form factors 
at the two $\pi BB$ vertices.
Lee and Matsuyama carried out an analysis of the $NN \to NN\pi$ processes 
and pointed out \cite{lee:prc} that a soft form factor, 
such as $\Lambda<750$MeV, is preferable.
Recent analysis in the QCD sum rule 
also suggests a soft $\pi NN$ form factor of cutoff 
$\Lambda \simeq 800$[MeV]~\cite{mei:prc}.
From the quark substructure point of view, 
it is also natural that the cutoff $\Lambda$ is 
of order $\sim 1/R_h \sim 400 - 500$MeV.
Table~\ref{ffc} shows the comparison of the `` hard '' 
(DP, $\Lambda_{DP}(\pi) = 1300$MeV) and `` soft '' 
(DP, $\Lambda_{DP} = 800$MeV) 
pion-exchange form factors for the $\Lambda N \to NN$ decay rates.
We find that the soft form factor reduces the decay rates by about $40\%$, 
while the hard one changes only by $\le 10\%$.

The G form factor is related to the Gaussian quark wave function of the baryon.
Corresponding to the size parameter $b \simeq 0.5$fm, 
we employ $\Lambda_G(\pi) \simeq 680$MeV 
because the relation between the size and cutoff parameters 
is $\Lambda =\sqrt{3}/b$.

The above sensitivity to choice of the form factor is quite annoying 
because it is not easy to judge which of the form factors is the right one.
It should also be noted that the weak vertex form factor 
can be different from the strong one, 
although the pole dominance picture of the parity-conserving weak vertex 
leads to the identical form factor to the strong one.
In the meson exchange potential models of the nuclear force, 
there seems a tendency to choose a hard form factor
because the heavy meson exchanges 
are often important for spin-dependent forces.
On the other hand, the quark model approach to the short range nuclear force 
gives significant spin dependencies comparable to the heavy meson exchanges 
and therefore the meson exchanges can be cut off rather sharply 
with a soft form factor.
In the present approach to the weak $\Lambda N \to NN$ interaction, 
whose short range part is described by the direct quark mechanism, 
we thus may follow the quark model approach to the nuclear force 
and take the `` soft '' form factor for OPE potential as a standard.
And we assume that the heavy meson ($K$, $\eta$, $\rho$, $\omega$, $K^\ast$) 
exchanges have the similar DP form factors 
with the cutoff given by the J\"{u}lich potential~\cite{jul:npa}.
We see that this choice gives the total decay rate of $\Lambda$ 
in nuclear matter with the recent experimental data 
for heavy hypernuclear decays.

\subsection{Decay Rates}
The calculated decay rates of $\Lambda$ in nuclear matter 
for several models are listed in Table~\ref{tot}.
\begin{table}[tb]
\caption{Nonmesonic decay rates of $\Lambda$ in nuclear matter 
         (in units of $\Gamma_\Lambda$). 
         The ``all'' includes 
         $\pi$, $K$, $\eta$, $\rho$, $\omega$, and $K^\ast$ meson exchanges.
         Some recent experimental data for medium-heavy hypernuclei
         are also given.}
\begin{center}
\begin{tabular}{||cl||c|c|c|c|c||} 
\hline
 & & $total$ & $\Gamma_p$ & $\Gamma_n$ & ${\Gamma_n}/{\Gamma_p}$ & $PV/PC$ \\
\hline
$\pi$      &DP hard & 2.575 & 2.354 & 0.221 & 0.094 & 0.337 \\
           &DP soft & 1.850 & 1.702 & 0.148 & 0.087 & 0.282 \\
           &no-f.f. & 2.819 & 2.571 & 0.248 & 0.097 & 0.358 \\
\hline
$\pi$+K    &DP hard & 1.099 & 1.075 & 0.024 & 0.022 & 0.631 \\
           &DP soft & 0.695 & 0.674 & 0.021 & 0.031 & 0.632 \\
           &no-f.f. & 1.143 & 1.111 & 0.032 & 0.028 & 0.745 \\
\hline
all        &DP hard & 1.672 & 1.571 & 0.101 & 0.064 & 1.468 \\
           &DP soft & 1.270 & 1.152 & 0.117 & 0.101 & 1.537 \\
           &no-f.f. & 1.744 & 1.704 & 0.040 & 0.024 & 2.952 \\
\hline
DQ         &        & 0.418 & 0.202 & 0.216 & 1.071 & 6.759 \\
\hline
DQ+$\pi$   &DP hard & 3.609 & 2.950 & 0.658 & 0.223 & 0.856 \\
           &DP soft & 2.726 & 2.202 & 0.523 & 0.238 & 0.896 \\
\hline
DQ+$\pi$+K &DP hard & 1.766 & 1.495 & 0.271 & 0.181 & 1.602 \\
           &DP soft & 1.204 & 0.998 & 0.206 & 0.207 & 1.954 \\
\hline
DQ+all     &DP hard & 3.591 & 3.019 & 0.572 & 0.189 & 3.188 \\
           &DP soft & 1.884 & 1.522 & 0.361 & 0.237 & 3.587 \\
\hline
\hline
$\Gamma(^{12}_\Lambda C)$ & EXP~\cite{jjs:prc} 
     & 1.14 $\pm$ 0.2 & --- & --- & 1.33$^{+1.12}_{-0.81}$ & ---  \\
$\Gamma(^{12}_\Lambda C)$ & EXP~\cite{hno:prc} 
     & 0.89 $\pm$ 0.15 $\pm$ 0.03 & 0.31$^{+0.18}_{-0.11}$ & --- 
                        & 1.87 $\pm$ 0.59$^{+0.32}_{-1.00}$ & --- \\
$\Gamma(^{12}_\Lambda C)$ & EXP~\cite{pby:psg} 
     & 1.14 $\pm$ 0.08 & --- & --- & --- & --- \\
$\Gamma(^{28}_\Lambda Si)$ & EXP~\cite{pby:psg} 
     & 1.28 $\pm$ 0.08 & --- & --- & --- & --- \\
$\Gamma(_\Lambda Fe)$      & EXP~\cite{pby:psg} 
     & 1.22 $\pm$ 0.08 & --- & --- & --- & --- \\
\hline
\end{tabular}
\end{center}
\label{tot}
\end{table}
We show not only the results of soft DP form factor 
but also the result of hard DP form factor.
The cutoff parameters for the soft DP form factor 
and the hard DP form factor are given in Table~\ref{ffc}.

First of all, we can see that the OPE model predicts 
a large total decay rate in comparison with experiment 
for the heavy hypernuclei, even if we choose the softer cutoff.
The tensor dominance property leads to 
the small $\Gamma_n/\Gamma_p$ and $PV/PC$ ratio.

One sees that the kaon exchange contribution 
reduces $\Gamma_p$ by more than factor two.
This mainly comes from the cancellation in the channel $d_p$.
At same time, the kaon exchange contribution reduces 
$\Gamma_n$.
Therefore, the n/p ratio remains small.
The decay is dominated by the $J=1$ channels.

When we include $\eta$, $\rho$, $\omega$, and $K^\ast$ (`` all '') mesons,
both the proton and neutron induced decays increase 
but the n/p ratio is still small ($\simeq 0.1$).
It is interesting to see that the PV/PC ratio becomes large,
when we include heavier mesons.

One sees in Table~\ref{tot} that 
the magnitude of DQ itself is small compared to that of $\pi$ or $\pi+K$.
However, the DQ has a large $\Gamma_n$
and a large n/p ratio.
It is also shown that the DQ mechanism is dominated by 
the parity violating channels and thus produces a large $PV/PC$ ratio.
The characteristic behaviors of DQ 
will distinguish it from the other mechanisms.

Now we go to the results of our complete approach, i.e., 
meson exchange mechanism plus direct quark mechanism.
One sees that the pion exchange contribution strongly depends 
on the choice of the form factor as stated before.
The hard form factor does not suppress the potential at short distances 
and yields a large OPE contribution.
As we assume that 
the DQ has little contribution to the tensor channel, 
$\Gamma_p$ is still too large in the case of $DQ+\pi$.
Again, the $K$ exchange reduces the tensor amplitude
and thus $\Gamma_p$ is suppressed by a factor 2.
However $\Gamma_n$ is reduced, at the same time, 
which results in the n/p ratio $\simeq 0.2$.
This value seems too small compared to the experimental values
for the medium and heavy hypernuclei.

Contribution beyond the $\pi$ and $K$ mesons
does not improve the situation.
It is also questionable whether the DQ mechanism 
and the vector meson exchanges are independent and can be superposed.
The double counting problem for the vector mesons 
and DQ contribution in nuclear force is pointed out in ref \cite{yaz:ppn}.
We thus take the `` $DQ+\pi+K$ '' with the soft $\pi$ form factor
as our present best model for the nonmesonic $\Lambda$ decay.

\section{Light Hypernuclei}
The same $\Lambda N \to NN$ transition potential 
is applied to the study of the nonmesonic weak decays 
of light s-shell hypernuclei,
${^5_\Lambda He}$, ${^4_\Lambda He}$, and ${^4_\Lambda H}$.
We assume the harmonic oscillator shell model wave function
for the initial nucleons, 
while the $\Lambda$ single particle wave function
calculated
with a realistic $\Lambda$-nuclear potential, which has a repulsion 
at short distances. (See ref~\cite{iom:npa} for details.)
We further consider the short-range correlation,
which is multiplied to the relative two-body wave functions.
We employ the same correlation functions
as those used in the nuclear matter calculation.

The Gaussian $b$ parameters  are 1.358 [fm] for A=5 and 1.65 [fm] for A=4.
For the final state, we assume the correlated plane waves
for the outgoing two nucleons, and sum up all the residual states.

Here we present the results for the light hypernuclei 
and compare with the experimental one.
Table~\ref{lig} summarizes the results given by the superposition
of the ME and DQ mechanism.

\begin{table}[tb]
\caption{Nonmesonic decay rates (in units of $\Gamma_\Lambda$)
         of light hypernuclei. 
         The DP (soft) form factor is used for OPE.}
\begin{center}
\begin{tabular}{||cl|c|c|c|c||} 
\hline
 & & $total$ & $\Gamma_p$ & $\Gamma_n$ & ${\Gamma_n}/{\Gamma_p}$ \\
\hline
${^5_\Lambda He}$ &  $\pi$   & 0.740 & 0.654 & 0.087 & 0.133 \\
                  & $\pi+K$  & 0.350 & 0.331 & 0.028 & 0.055 \\
                  &$\pi+K+DQ$& 0.521 & 0.435 & 0.085 & 0.195 \\
\cline{3-6}
                  & EXP~\cite{jjs:prc} & 0.41$\pm$0.14 & 0.21$\pm$0.07 &
                               0.20$\pm$0.11 & 0.93$\pm$0.55 \\
                  & EXP~\cite{hno:psg} & 0.50$\pm$0.07 & 0.17$\pm$0.04 &
                               0.33$\pm$0.04 & 1.97$\pm$0.67 \\
 \hline
\hline
${^4_\Lambda He}$ &  $\pi$   & 0.542 & 0.498 & 0.044 & 0.089 \\
                  & $\pi+K$  & 0.252 & 0.233 & 0.019 & 0.082 \\
                  &$\pi+K+DQ$& 0.309 & 0.302 & 0.007 & 0.024 \\
\cline{3-6}
                  & EXP~\cite{hno:psg} & 0.19$\pm$0.04 & 0.15$\pm$0.02 
                             & 0.04$\pm$0.02 & 0.27$\pm$0.14 \\
 \hline
\hline
${^4_\Lambda H}$  &  $\pi$   & 0.080 & 0.022 & 0.056 & 2.596 \\
                  & $\pi+K$  & 0.020 & 0.010 & 0.010 & 1.099 \\
                  &$\pi+K+DQ$& 0.120 & 0.060 & 0.059 & 0.983 \\
\cline{3-6}
                  & EXP~\cite{hno:psg} & 0.15$\pm$0.13 & ----- 
                                             & ----- & ----- \\
 \hline
\hline
\end{tabular}
\end{center}
\label{lig}
\end{table}

In this result we find that the ``$\pi$ + $K$ + DQ'' picture 
gives us good agreement with experiment for the total decay rate,
while the n/p ratios are still too small.
As seen in the nuclear matter calculation, 
the main contribution comes from the OPE mechanism, 
which produces a large proton-induced rate.
Comparing with the experimental data 
we find that the proton-induced rate is overestimated in all the pictures, 
while the neutron-induced rate is underestimated.

\section{Summary and Conclusion}
In this paper we study the $\Lambda N \to NN$ weak transition
by combining the ME and DQ mechanisms.
We use the weak meson-baryon coupling vertices 
which are evaluated by using $SU(6)_w$ symmetry.
The DQ mechanism comes from the quark structure of two baryons 
and is effective at the short range 
where two baryons overlap with each other.
Adding these two types of contributions,
we calculate the decay rate of $\Lambda$ in nuclear matter 
and in light hypernuclei.
The choice of the form factor for the $\pi$ exchange
induced potential is found to be important.
With the heavier mesons exchange 
or DQ for the short range part,
it seems appropriate to employ the soft form factor for OPE.

In our calculation we find that the choice of `` $\pi + K + DQ$ '' 
with soft DP form factor for OPE 
reproduces the current experimental data rather well.
The OPE contribution dominates the decay.
The $K$ contribution is also large 
and interferes destructively with the $\pi$ contribution.
These two contributions yield the long range part of this process.
In the `` $\pi + K + DQ$ '' picture 
it is assumed that the DQ mechanism 
replaces the role of the heavier mesons.

Although the present analysis gives the total decay rates
both in nuclear matter and in light hypernuclei fairly well,
there remains a difficult problem.
That is, the proton-induced decay rate is too large
compared to experimental one and thus we predict a small n/p ratio.
The ratio is improved from the $\pi+K$ exchange prediction
due to the DQ contribution,
in which the n/p ratio is about unity.
It is, however, still small ($\simeq 0.2$) for `` $\pi + K + DQ$ ''.
The experimental numbers are not completely fixed,
but they suggest the value around $1$ for heavy hypernuclei.
The situation is similar for $^5_\Lambda$He.
It is thus urgent and important to find out 
what causes this discrepancy.

\section*{Appendix}

In this Appendix, we present the explicit forms of the DQ transition 
potential in the coordinate space.  This is based on its momentum 
representation given in ref.~\cite{iok:npa} (referred to as (I) hereafter). 
The potential is expressed in terms of three sets of seven functions 
of $k$ and $k'$, 
$f(k,k')_{i}$, $g(k,k')_{i}$, and $h(k,k')_{i}$, given 
in Tables 6, 7, and 8 of the paper (I).
The complete transition potential is given by eq.(24) of the paper (I)
as a combination of the functions, $f$, $g$, and $h$'s, and the 
coefficients, which depends on the initial and final orbital 
angular momenta, spins and isospin.
Those coefficients denoted by $V_{i}$'s and $W$ are 
explicitly given in Tables 2, 3, 4, and 5 in the paper (I), 
so that anyone can reconstruct the whole transition 
potential without any further information.

The purpose of this Appendix is to replace the functions, 
$f$, $g$, and $h$'s, by their Fourier transformed counterparts, which 
are convenient for hypernuclear decays with sophisticated nuclear 
wave functions.  
In the coordinate space, the potential becomes nonlocal due to the 
quark antisymmetrization effect, and also contains a derivative term, 
which generates a transition from $L=0$ to $L'=1$.  Such derivative 
terms appear for the functions, $g$ and $h$'s.
The complete potential is now given as a nonlocal form as
\begin{equation}
   {V_{DQ}}^{L L'}_{S S'J}(r,r')
    =  -{G_{F}\over\sqrt{2}} \times W \, \sum_{i=1}^{7}
    \left\{ V_{i}^f f_{i}(r, r') + V_{i}^g g_{i}(r, r') 
    + V_{i}^h h_{i}(r, r') 
    \right\}
\end{equation}
where $r$ ($r'$) stands for the radial part of the relative coordinate 
in the initial (final) state.
We only need to give the forms of the radial functions, $f$, $g$, 
and $h$'s as the coefficients, $V_{i}$'s and $W$, are the same 
ones as in the paper (I).  
(We, however, call the reader's attention that the 
transition potential given in the paper (I) is for the antisymmetrized 
initial states.  They are antisymmetrized so that the flavor 
antisymmetric states are defined as $(p\Lambda-\Lambda p)/\sqrt{2}$
or $(n\Lambda-\Lambda n)/\sqrt{2}$.  Thus they have an opposite sign 
to those commonly used in the literature for the meson exchange 
potential. It is necessary to take care of this difference in 
convention when we superpose the DQ potential with that from 
the meson exchange.  If one adds our transition potential to the OPE 
in the conventional definition, we need to change the signs of the 
coefficients for the flavor-antisymmetric initial states.)

The local parts (with a derivative on the initial radial coordinate) 
are given for $i=1$, 6, and 7, by
\begin{eqnarray}
    f_{1}(r, r') & =& {6\over N} 1\cdot 1\cdot (2\pi b^2)^{-3/2} 
    {\delta(r-r')\over r^2}
    \nonumber  \\
    g_{1}(r, r') & =& {6\over N} 1\cdot {1\over\sqrt{3}m} 
    \cdot (2\pi b^2)^{-3/2} (-i) {\delta(r-r')\over r^2} \partial_{r}
      \\
    h_{1}(r, r') & =& {6\over N} 1\cdot {1\over\sqrt{3}m} 
    \cdot (2\pi b^2)^{-3/2} (-i) {\delta(r-r')\over r^2} \partial_{r}
    \nonumber 
\\ \bigskip
    f_{6}(r, r') & =& {9\over N} 1\cdot 1\cdot (4\pi A)^{-3/2} 
    e^{-r^2/4A} {\delta(r-r')\over r^2}
    \nonumber  \\
    g_{6}(r, r') & =& {9\over N} 1\cdot {1\over\sqrt{3}m} 
    \cdot (4\pi A)^{-3/2} e^{-r^2/4A} 
    (-i) {\delta(r-r')\over r^2} \partial_{r}
      \\
    h_{6}(r, r') & =& {9\over N} 1\cdot {1\over\sqrt{3}m} 
    \cdot (4\pi A)^{-3/2} e^{-r^2/4A} 
     (-i) {\delta(r-r')\over r^2} \left[\partial_{r} + {1\over 2A} r 
     \right]
    \nonumber
\\ \bigskip
    f_{7}(r, r') & = & -{1\over 3} f_{6}(r, r') 
    \nonumber\\
    g_{7}(r, r') & = & -{1\over 3} g_{6}(r, r') 
      \\
    h_{7}(r, r') & = & -{1\over 3} h_{6}(r, r') 
    \nonumber
\end{eqnarray}
where $A\equiv b^2/3$, the normalization factor $N$ is taken as $N=1$ and 
$\partial_{r}$ is the derivative on the initial radial coordinate.

Among the above local potentials, the $f_{1}$ part contains a 
constant term, which survives at $r=r' \to\infty$.  This term comes 
from the internal weak conversion of $\Lambda$ into a neutron and is
to be suppressed due to the mismatch of the initial and final wave 
functions.  However, in the application to hypernuclear decays we do 
not necessarily employ wave functions which satisfy the orthogonality 
condition of the bound and scattering states.  Thus the $f_{1}$ term 
may survive in the actual calculation.  As this term is spurious in 
most of the cases, we omit the $f_{1}$ term in the actual calculation.

The nonlocal parts are given in terms of the functions,
\begin{eqnarray}
    G_{0} & \equiv & (4\pi^2 D)^{-3/2} 4\pi i_{0}\left({Crr'\over 
    D}\right) \exp\left({ -B'r'^2-Br^2\over D}\right)
    \nonumber\\
    G_{1} & \equiv & (4\pi^2 D)^{-3/2} 4\pi i_{1}\left({Crr'\over 
    D}\right) \exp\left({ -B'r'^2-Br^2\over D}\right)
\end{eqnarray}
where $i_{\ell}$ is the $\ell$th modified spherical Bessel function.

For $i=2$, the constants $B$, $B'$, $C$, and $D$ are given by
\begin{equation}
    B=B' =  {5b^2\over 12} \quad
    C  =  {b^2\over 2} \quad
    D  =  {4 b^4\over 9}
\end{equation}
and we find
\begin{eqnarray}
    f_{2}(r, r') & =& -{18\over N}\,{1\over 3}\cdot 1\cdot 
    {3\sqrt{6}\over 4} G_{0} 
    \nonumber  \\
    g_{2}(r, r') & =& -{18\over N}\,{1\over 3}\cdot {1\over\sqrt{3}m} 
    \cdot {3\sqrt{6}\over 4} 
    (i) \left[ -{2B\over D} r G_{1}+ {C\over D} r'G_{0} \right]
      \\
    h_{2}(r, r') & =& -{18\over N}\,{1\over 3}\cdot {1\over\sqrt{3}m} 
    \cdot {3\sqrt{6}\over 4}
    (-i) \left[ -{2B'\over D} r' G_{0}+ {C\over D} rG_{1} \right]
    \nonumber
\end{eqnarray}
For $i=4$,
\begin{eqnarray}
    f_{4}(r, r') & =& -{36\over N}\,{1\over 3}\cdot 1\cdot 
    {3\sqrt{3}\over 8}  G_{0} 
    \nonumber  \\
    g_{4}(r, r') & =& -{36\over N}\,{1\over 3}\cdot {1\over\sqrt{3}m} 
    \cdot {3\sqrt{3}\over 8} 
    (i) \left[ -{2B\over D} r G_{1}+ {C\over D} r'G_{0} \right]
      \\
    h_{4}(r, r') & =& -{36\over N}\,{1\over 3}\cdot {1\over\sqrt{3}m} 
    \cdot {3\sqrt{3}\over 8} 
    (-i) \left[ -{2B'\over D} r' G_{0}+ {C\over D} rG_{1} \right]
    \nonumber
\end{eqnarray}
with 
\begin{equation}
    B=B'  =  {b^2\over 6} \quad
    C =  0\quad
    D =  {b^4\over 9}
\end{equation}
For $i=3$, and $i=5$, we find
\begin{eqnarray}
    f_{3}(r, r') & =& -{36\over N}\,{1\over 3}\cdot 1\cdot 
    {24\sqrt{33}\over 121}  G_{0} 
    \nonumber  \\
    g_{3}(r, r') & =& -{36\over N}\,{1\over 3}\cdot {1\over\sqrt{3}m} 
    \cdot {24\sqrt{33}\over 121} 
    (i) \left[ -{2B\over D} r G_{1}+ {C\over D} r'G_{0} \right]
      \\
    h_{3}(r, r') & =& -{36\over N}\,{1\over 3}\cdot {1\over\sqrt{3}m} 
    \cdot {24\sqrt{33}\over 121} 
    (-i) \left[ -{2B'\over D} r' G_{0}+ {C\over D} rG_{1} \right]
    \nonumber
\end{eqnarray}
with for $i=3$,
\begin{equation}
    B  =  {13b^2\over 33}\quad
    B'  =  {7b^2\over 33}\quad
    C  =  {12b^2\over 33}\quad
    D  =  {20b^4\over 99}
\end{equation}
and for $i=5$,
\begin{equation}
    B  =  {7b^2\over 33} \quad
    B'  =  {13b^2\over 33} \quad
    C  =  {12b^2\over 33} \quad
    D  =  {20b^4\over 99}
\end{equation}

\begin{figure}[bt]
        \centerline{ \epsfxsize=7.5cm \epsfbox{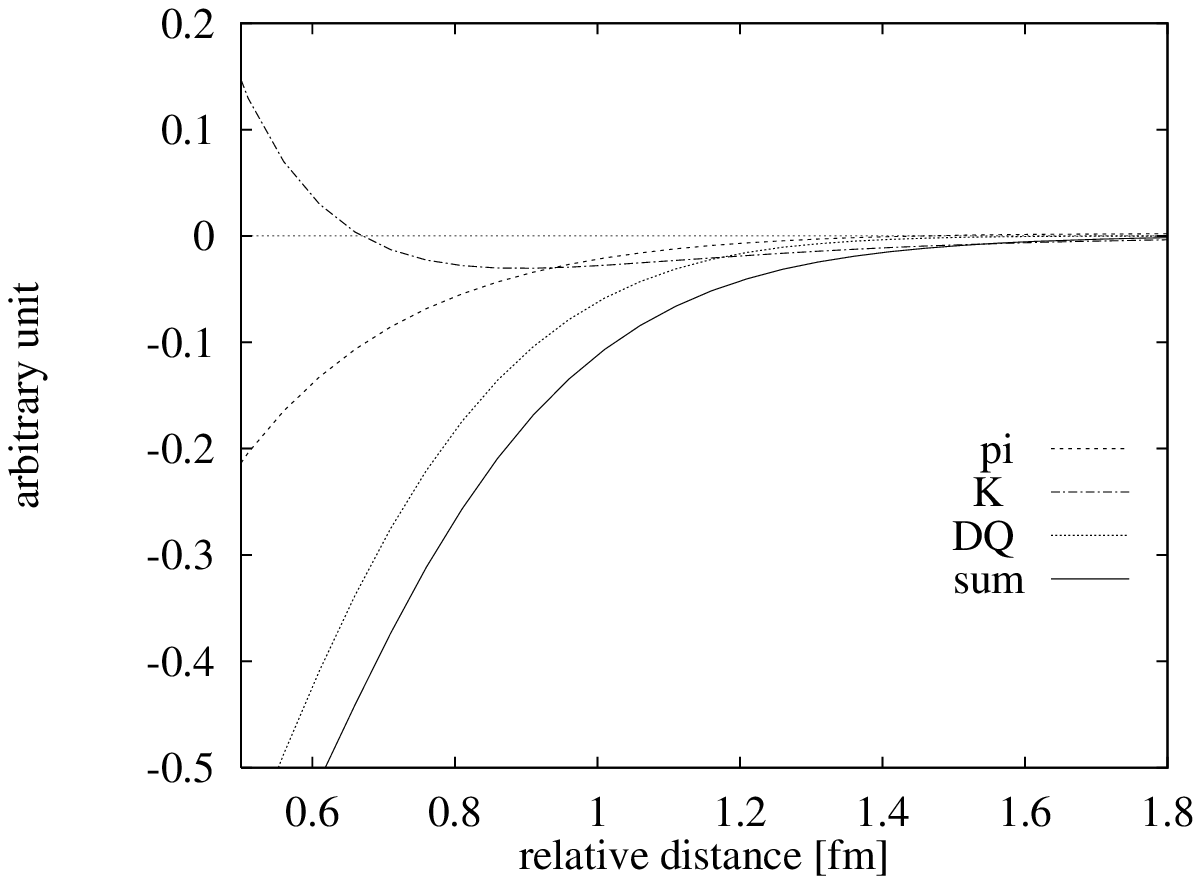} ~\ ~\
                     \epsfxsize=7.5cm \epsfbox{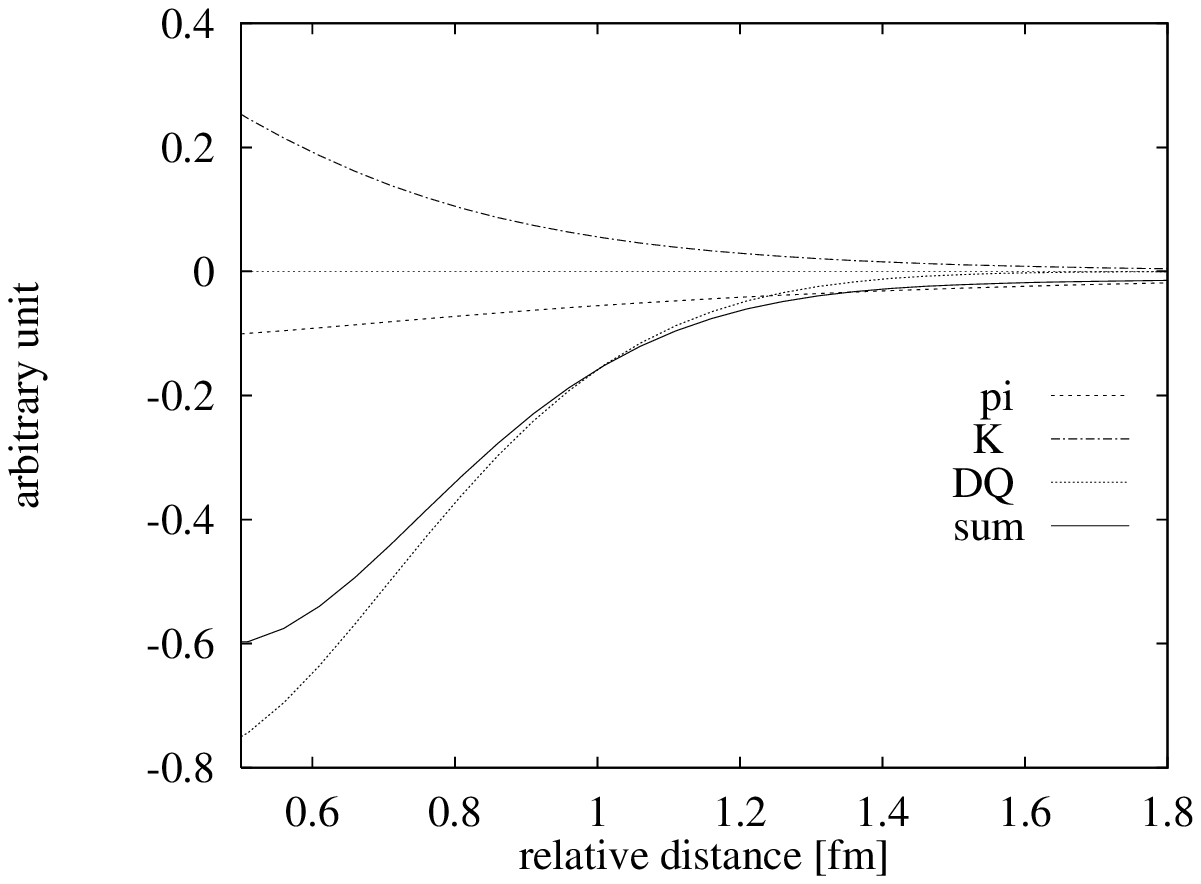} } \vspace{-0.7cm}
        \begin{center} $a_p$ \hspace{7.5cm} $b_p$  \end{center}
        \vspace{0.1cm}
        \centerline{ \epsfxsize=7.5cm \epsfbox{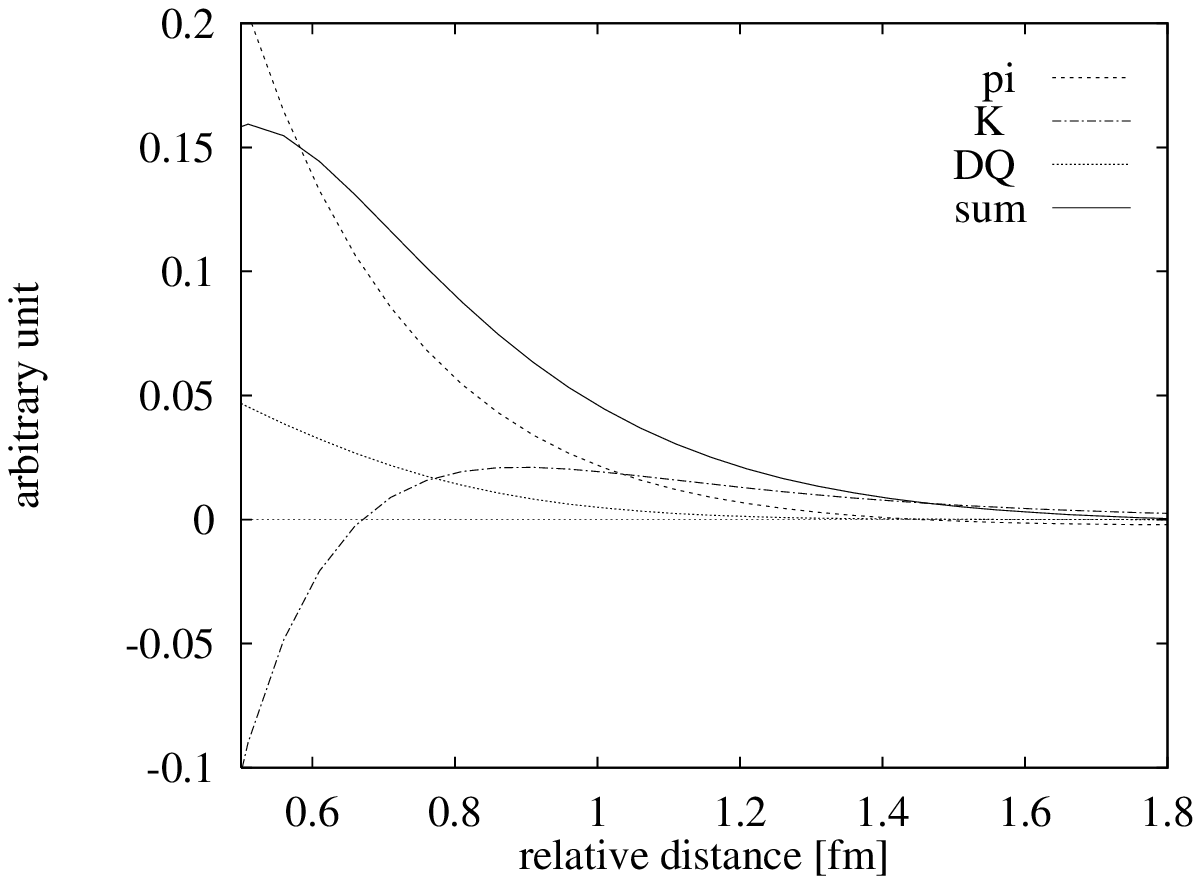} ~\ ~\
                     \epsfxsize=7.5cm \epsfbox{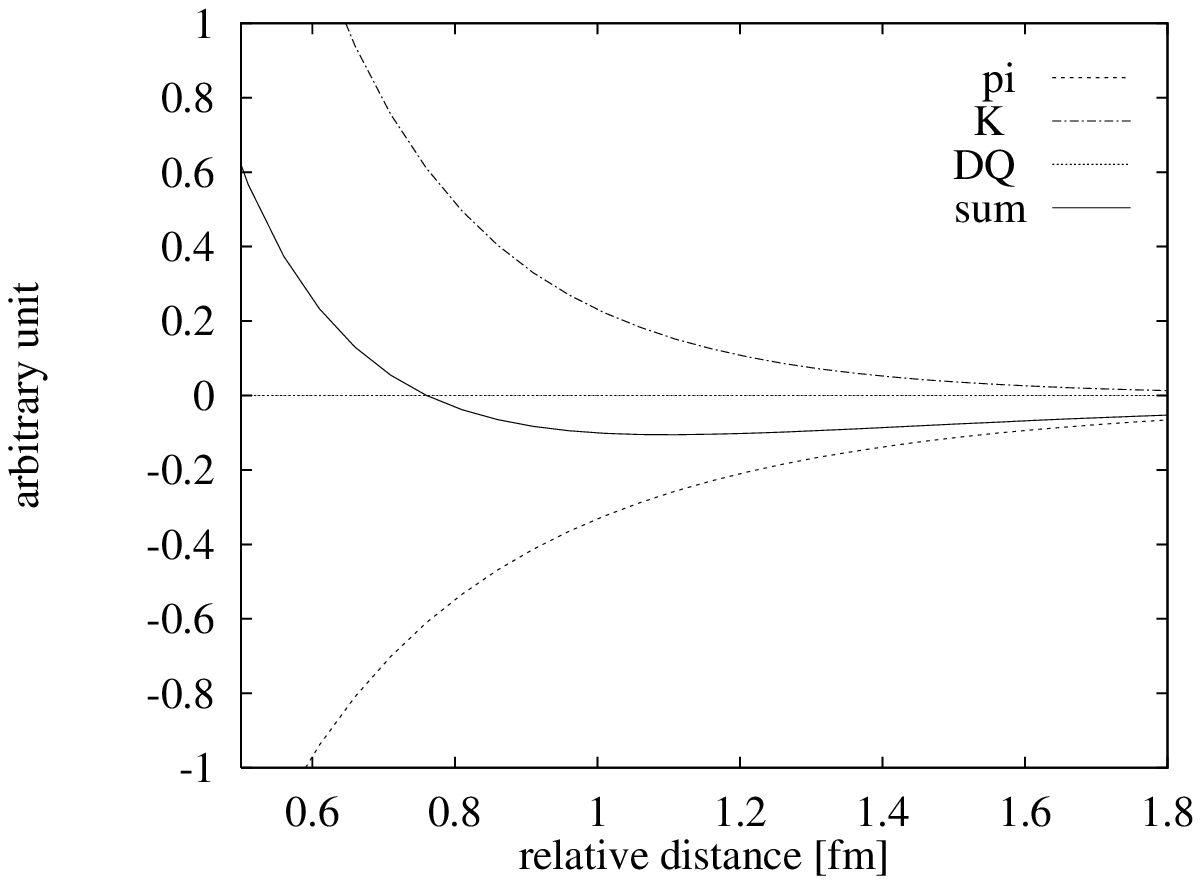} } \vspace{-0.7cm}
        \begin{center} $c_p$ \hspace{7.5cm} $d_p$  \end{center}
        \vspace{0.1cm}
        \centerline{ \epsfxsize=7.5cm \epsfbox{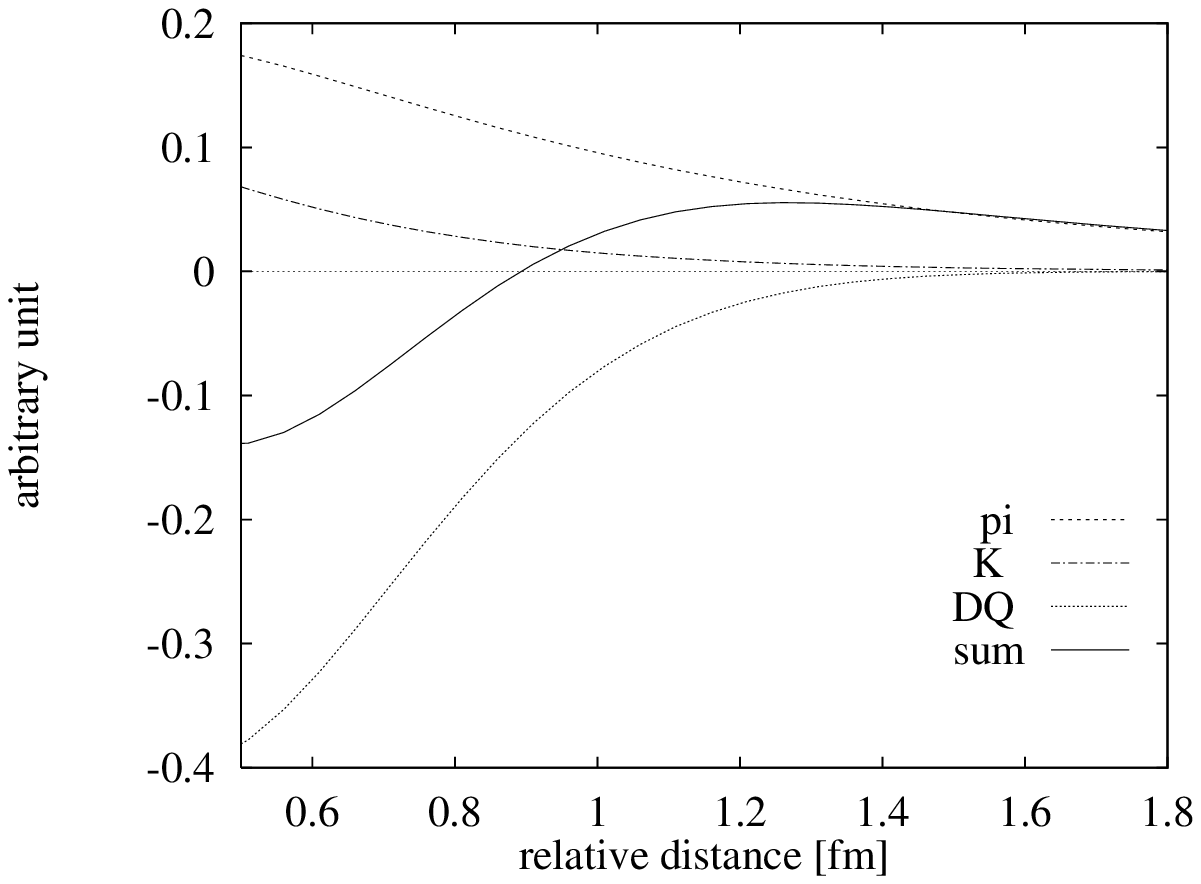} ~\ ~\
                     \epsfxsize=7.5cm \epsfbox{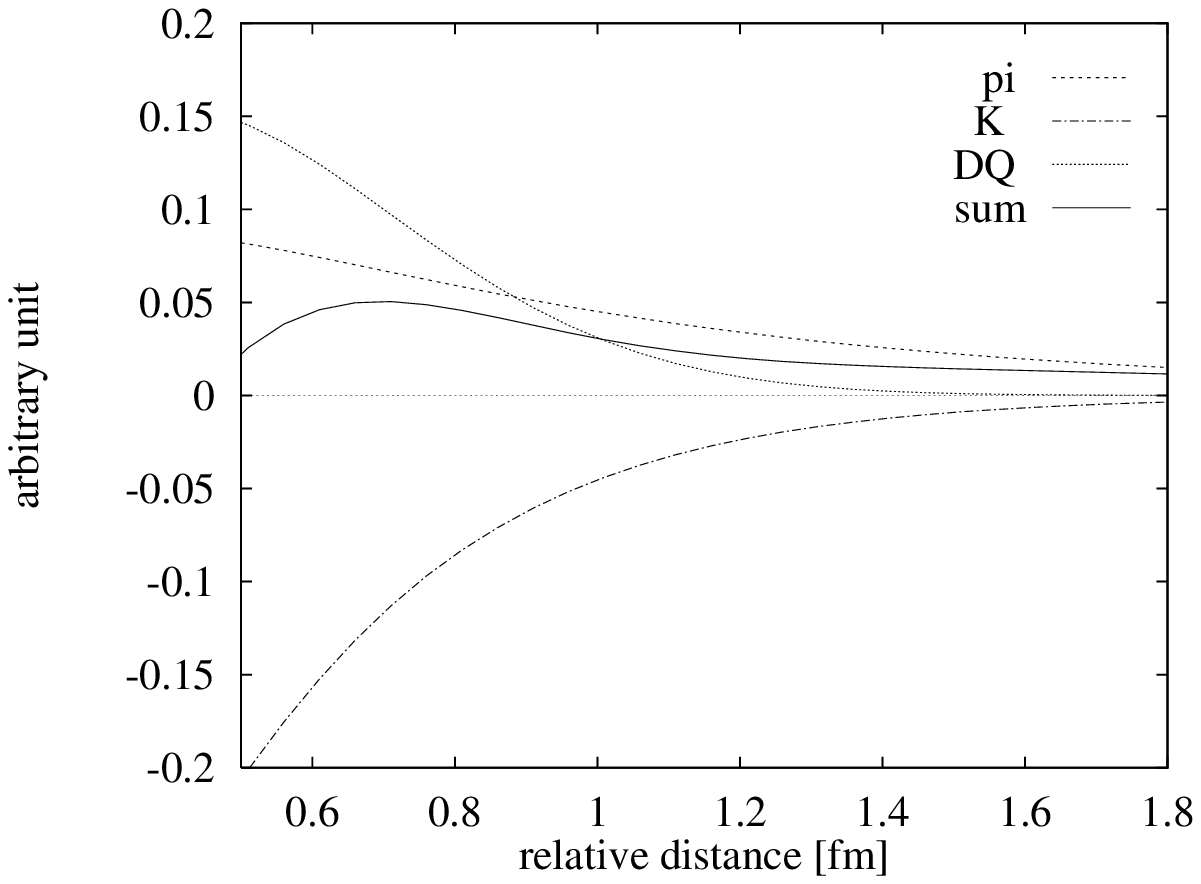} } \vspace{-0.7cm}
        \begin{center} $e_p$ \hspace{7.5cm} $f_p$  \end{center}
        \vspace{-0.3cm}
        \caption{The transition potentials for the proton-induced decay.} 
\label{pid}
\end{figure}

\begin{figure}[bt]
        \centerline{ \epsfxsize=7.5cm \epsfbox{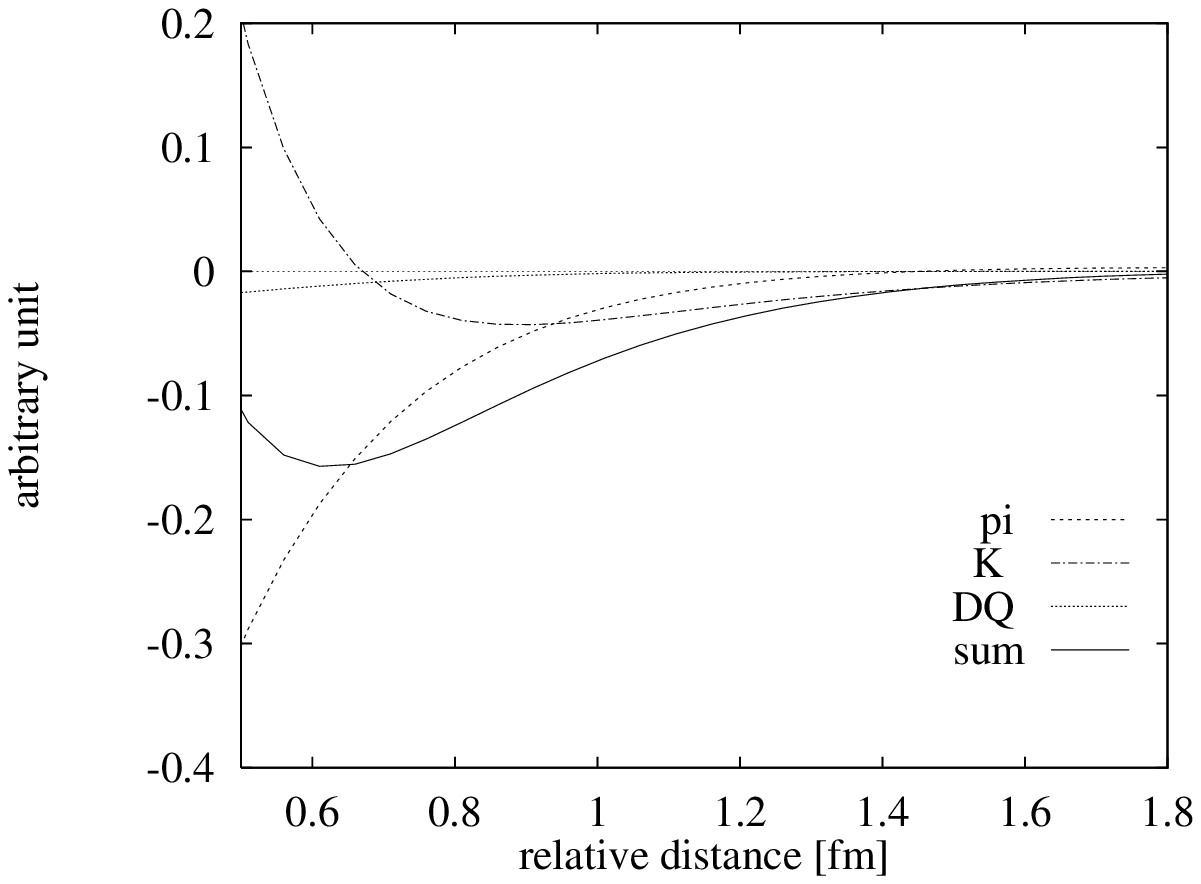} ~\ ~\
                     \epsfxsize=7.5cm \epsfbox{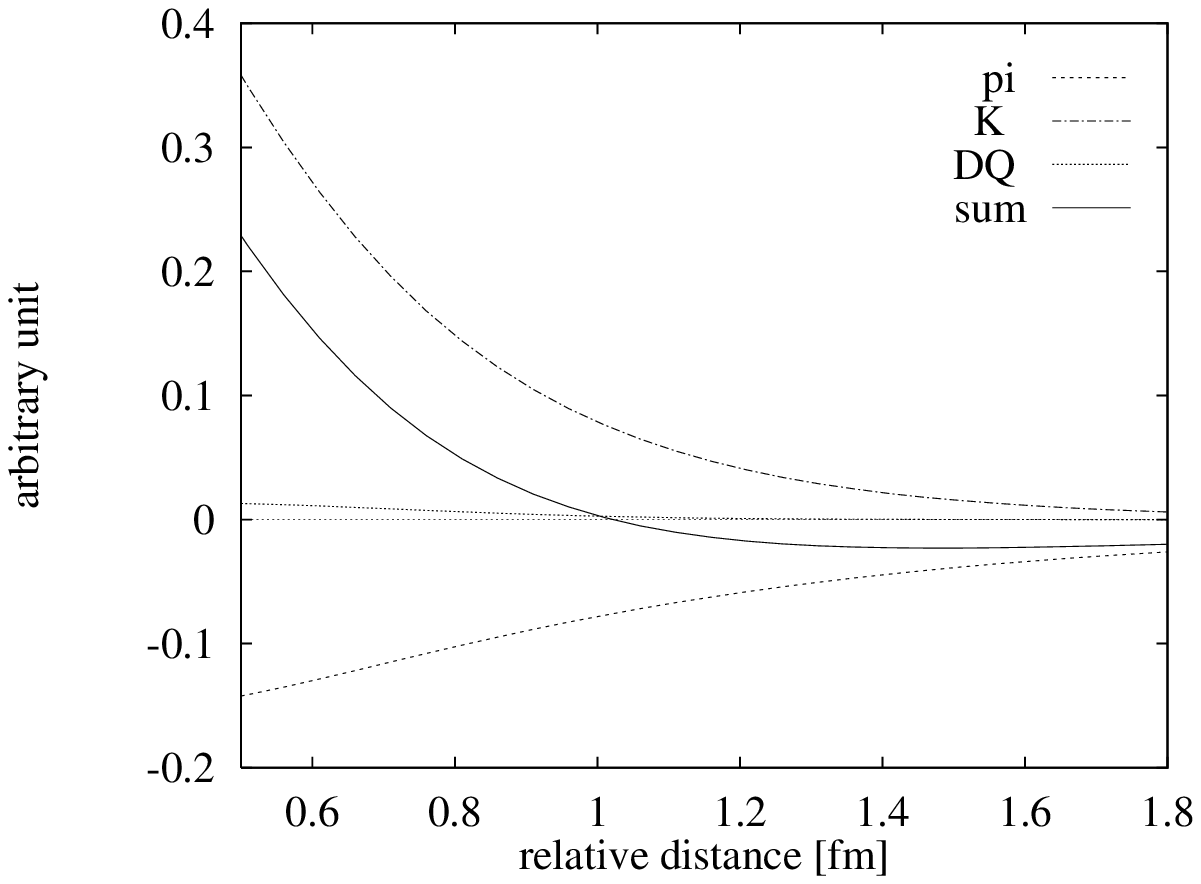} } \vspace{-0.7cm}
        \begin{center} $a_n$ \hspace{7.5cm} $b_n$  \end{center}
        \vspace{0.1cm}
        \centerline{ \epsfxsize=7.5cm \epsfbox{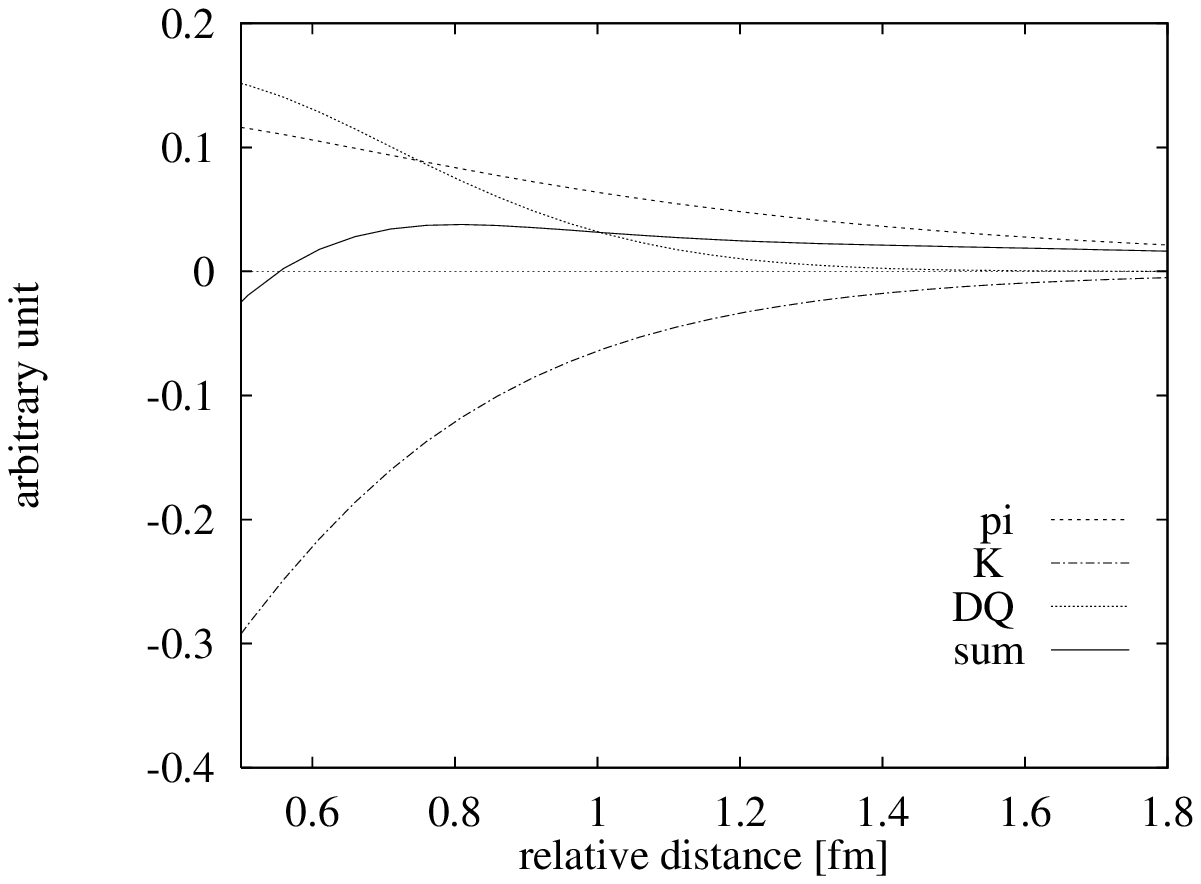} ~\ ~\
                     \hspace{7.5cm}} \vspace{-0.7cm}
        \begin{center} $f_p$ \hspace{7.5cm} {}     \end{center}
        \vspace{-0.3cm}
        \caption{The transition potentials for the neutron-induced decay.} 
\label{nid}
\end{figure}

The local parts of the DQ, $\pi$-, $K$- exchange potentials
are illustrated in Fig.~\ref{pid} for the proton induced transitions
and Fig.~\ref{nid} for the neutron induced one.
The meson exchange potentials are multiplied by the form factors.
We take the soft ($\Lambda_\pi = 800$MeV) DP form factor for OPE.


\newpage


\begin{thebibliography}{99}
\bibitem{dgh:pre} 
    J. F. Donoghue, E. Golowich, and B. Holstein,  
    Phys. Rep. ${\bf{131}}$ 319 (1986).
\bibitem{bld:prl} 
    M. M. Block, and R. H. Dalitz,
    Phys. Rev. Lett. ${\bf{11}}$ 96 (1963).
\bibitem{mcg:prc} 
    B. H. J. McKeller, and B. F. Gibson, 
    Phys. Rev. C${\bf{30}}$, 322 (1984).
\bibitem{ttb:ptp} 
    K. Takeuchi, H. Takaki, and H. Bando, 
    Prog. Theor. Phys. ${\bf{73}}$ 841 (1985);
    H. Bando, Y. Shono, and H. Takaki,
    Int. Jour. Mod. Phys. A${\bf{3}}$ 1581 (1988).
\bibitem{ost:npa} 
    E. Oset and L. L. Salcedo, 
    Nucl. Phys. A${\bf{443}}$ 704 (1985);
    A. Ramos, E. Oset, and L. L. Salcedo,
    Phys. Rev. C${\bf{50}}$ 2314 (1994).
\bibitem{bmz:jmp} 
    H. Bando, T. Motoba, and J. Zofka,
    Int. Jour. Mod. Phys. A${\bf{5}}$ 4021 (1990).
\bibitem{rmb:npa} 
    A. Ramos, E. van Meijgaard, C. Bennhold, and B. K. Jennings,
    Nucl. Phys. A${\bf{544}}$ 703 (1992).
\bibitem{prb:prc} 
    A. Parre$\tilde{\rm{n}}$o, A. Ramos, and E. Oset, 
    Phys. Rev. C${\bf{51}}$, 2479 (1995);
    A. Parre$\tilde{\rm{n}}$o, A. Ramos, and C. Bennhold, 
    Phys. Rev. C${\bf{52}}$, R1768 (1995).
\bibitem{dub:anp} 
    J. F. Dubach, G. B. Feldman, B. R. Holstein, and L. de la Torre, 
    Ann. Phys. ${\bf{249}}$, 146 (1996).
\bibitem{par:prc} 
    A. Parre$\tilde{\rm{n}}$o, A. Ramos, and C. Bennhold, 
    Phys. Rev. C${\bf{56}}$, 339 (1997).
\bibitem{chk:prc} 
    C. H. Cheung, D. P. Heddle, and L. S. Kisslinger, 
    Phys. Rev. C${\bf{27}}$ 335 (1983).
\bibitem{mal:plb} 
    K. Maltman and M. Shmatikov,  
    Phys. Lett. B${\bf{331}}$ 1(1994).
\bibitem{iok:npa} 
    T. Inoue, S. Takeuchi, and M. Oka, 
    Nucl. Phys. A${\bf{577}}$ 281c (1994);
    {\it{ibid}} A${\bf{597}}$ 563 (1996).
\bibitem{iom:npa} 
    T. Inoue, M. Oka, T. Motoba, and K. Itonaga, 
    Nucl. Phys. A${\bf{633}}$ 312 (1998).
\bibitem{shm:npa} 
    M. Shmatikov,
    Nucl. Phys. A${\bf{580}}$ 538 (1994).
\bibitem{ium:npa} 
    K. Itonaga, T. Ueda, and T. Motoba,
    Nucl. Phys. A${\bf{639}}$ 329c (1998).
\bibitem{jjs:prc} 
    J. J. Szymznski et al., 
    Phys. Rev. C${\bf{43}}$ 849 (1991).
\bibitem{hno:prc} 
    H. Noumi et al., 
    Phys. Rev. C${\bf{52}}$ 2936 (1995).
\bibitem{pby:psg} 
    H. Bhang et al,. 
    Phys. Rev. Lett. ${\bf{81}}$ 4321 (1998).
\bibitem{hno:psg}
    H. Noumi et al,. 
    in proceedings of the IV International Symposium
         on Weak and Electromagnetic Interactions in Nuclei,
         edited by H. Ejiri, T. Kishimoto and T. Sato 
         (World Scientific,1995) p.550
\bibitem{nrs:prd} 
    M. N. Nagels, T. A. Rijken, and J. J. Swart,
    Phys. Rev. D${\bf{15}}$, 2547 (1977);
    P. M. M. Maessen, T. A. Rijken, and J. J. de Swart,
    Phys. Rev. C${\bf{40}}$, 2226 (1989).
\bibitem{vgj:prc}
     V. G. J. Stoks, R. A. M. Klomp, C. P. F. Terheggen, and J. J. de Swart,
     Phys. Rev. C{\bf{49}} 2950 (1994).
\bibitem{QCM}
    M. Oka and K. Yazaki, 
    Phys. Lett. B${\bf{90}}$ 41 (1980);
    Prog. Theor. Phys. ${\bf{66}}$ 556 (1981)
    {\it{ibid}} ${\bf{66}}$ 572 (1981);
         in {\sl Quarks and Nuclei}, 
         ed. by W. Weise (World Scientific, 1985);
    K. Shimizu, 
    Rep. Prog. Phys. ${\bf{52}}$ 1 (1989).
\bibitem{psw:npb}
    M. K. Gaillard and B. W. Lee, 
    Phys. Rev. Lett. ${\bf{33}}$ 108 (1974);
    A. I. Vainshtein, V. I. Zakharov and M. A. Shifman, 
    Sov. Phys. JETP ${\bf{45}}$ 670 (1977);
    F. J. Gillman, M. B. Wise, 
    Phys. Rev. D${\bf{20}}$ 2382 (1979);
    E. A. Paschos, T. Schneider and Y. L. Wu, 
    Nucl. Phys. B${\bf{332}}$ 285 (1990).

\bibitem{jul:npa}
     B. Holzenkamp, K. Holinde, and J. Speth,
     Nucl. Phys. A{\bf{500}} 485 (1989).
\bibitem{mhe:pre}
    R. Machleidt, K. Holinde, and Ch. Elster,
    Phys. Rep. ${\bf{149}}$, 1 (1987)
\bibitem{lee:prc} 
    T. S. H. Lee and A. Matsuyama, 
    Phys. Rev. C${\bf{36}}$ 1459 (1987).
\bibitem{mei:prc}
     T. Meissner,
     Phys. Rev. C{\bf{52}} 3386 (1995).
\bibitem{yaz:ppn} 
    K. Yazaki, 
    Prog. Part. Nucl. Phys. ${\bf{24}}$ 353 (1990).
\end{thebibliography}
\end{document}